\documentclass[lengthcheck,onecolumn,pre,superscriptaddress,showpacs,floatfix,aps]{revtex4-2}

\usepackage{amsmath,amssymb,amsfonts,latexsym,bm}
\usepackage{graphicx,epsfig,subcaption}
\usepackage{fancyvrb}
\usepackage{booktabs}
\usepackage{soul}
\usepackage{algorithm}
\usepackage{algpseudocode}

\usepackage{ragged2e}
\usepackage[svgnames]{xcolor}
\usepackage{xfrac}
\usepackage{mathrsfs}

\usepackage[colorlinks,linkcolor=MediumBlue,citecolor=MediumBlue,urlcolor = MediumBlue]{hyperref}
\usepackage{url}

\allowdisplaybreaks

\begin{document}

\title{Many wrong models approach to localize an odor source in turbulence with static sensors} 

\author{Lorenzo Piro}\email{lorenzo.piro@roma2.infn.it}
\affiliation{Department of Physics \& INFN, University of Rome ``Tor Vergata", Via della Ricerca Scientifica 1, 00133 Rome, Italy}
\author{Robin A. Heinonen}
\affiliation{Department of Physics \& INFN, University of Rome ``Tor Vergata", Via della Ricerca Scientifica 1, 00133 Rome, Italy}
\author{Massimo Cencini}
\affiliation{Istituto dei Sistemi Complessi, CNR, Via dei Taurini 19, 00185 Rome, Italy}
\affiliation{INFN ``Tor Vergata", Via della Ricerca Scientifica 1, 00133 Rome, Italy}
\author{Luca Biferale}
\affiliation{Department of Physics \& INFN, University of Rome ``Tor Vergata", Via della Ricerca Scientifica 1, 00133 Rome, Italy}

\date{\today}

\begin{abstract}
    The problem of locating an odor source in turbulent flows is central to key applications such as environmental monitoring and disaster response. We address this challenge by designing an algorithm based on Bayesian inference, which uses odor measurements from an ensemble of static sensors to estimate the source position through a stochastic model of the environment. 
    The problem is difficult because of the multiscale and out-of-equilibrium properties of turbulent transport, which lack accurate analytical and phenomenological modeling, thus preventing a guaranteed convergence for Bayesian approaches. To overcome the risk of relying on a single unavoidably wrong model approximation, we propose a method to rank ``many wrong models'' and to blend their predictions. We evaluated our \emph{weighted Bayesian update} algorithm by its ability to estimate the source location with predefined accuracy and/or within a specified time frame and compare it to standard Monte Carlo sampling methods. To demonstrate the robustness and potential applications of both approaches under realistic environmental conditions, we use high-quality direct numerical simulations of the Navier-Stokes equations to mimic the turbulent transport of odors in presence of a strong mean wind. Despite minimal prior information on the source and environmental conditions, our proposed approach consistently proves to be more accurate, reliable, and robust than Monte Carlo methods, thus showing promise as a new tool for addressing the odor source localization problem in real-world scenarios.
\end{abstract}

\maketitle

\noindent{\textbf{Keywords:} Bayesian inference; Monte Carlo methods;  turbulence; source term estimation; many wrongs principle; olfactory search; sensors.

\section{Introduction}

Identifying the origin of noxious odors such as gas leaks and chemical or radioactive emissions is critical for averting potential environmental disasters~\cite{bayat2017_review,burgues2020_review,francis2022_review,Karafasoulis2023_radio}. Moreover, tracing the location of a source from the odor signal detected in the atmosphere is a vital task for animals looking for food or a mate~\cite{murlis1992_review,hein2016}.

Such source localization problems are already difficult in laminar flows, where the structure of the odor plume can be highly complex and sensitive to the source location due to chaotic mixing~\cite{aref2017,speetjens2021}. In fully turbulent three-dimensional (3D) environments, the difficulty becomes severe due to numerical, experimental, and phenomenological complexities and challenges associated with the multiscale nature of 3-D turbulence, which is characterized by a wide range of rough velocity fluctuations, as well as highly intermittent and non-Gaussian properties for both advecting flow and odor transport~\cite{frisch1995,eyink1996,alexakis2018,crimaldi2001,crimaldi2002,celani2014}. 
%The primary challenge stems from the fact that turbulent dispersion results in the evolution of the odor signal being unpredictable and spatiotemporally intermittent~\cite{crimaldi2001,celani2014}. 

This makes the design of efficient and reliable algorithms for source localization a challenging theoretical task located at the intersection of fluid dynamics, optimal mobile agent navigation, and information theory~\cite{celani2014,piro2023_thesis,reddy2022_review}.
In recent decades, a growing community of scientists has developed multiple methodologies to address this problem, ranging from bioinspired~\cite{belanger1998biologically,balkovsky2002,durve2020} to heuristic search algorithms based on information theory~\cite{vergassola2007,masson2009,loisy2022} and Bayesian inference~\cite{keats2007,hutchinson2019}.

Although most studies have focused on optimizing mobile agents' performance, the use of static sensors has become increasingly popular~\cite{aslam2003,shankarrao2007,hutchinson2017_review} due to the simplicity of their implementation, which also results in reduced setup and maintenance costs.
In addition, being well suited for continuous and large-scale environmental monitoring, they are ideal for use in early warning systems~\cite{sattele2015,stahli2015,tariq2021,esposito2022}, which may enable targeted and effective response strategies, ranging from containment and mitigation to rescue and prevention. 
On a more theoretical level, the use of static sensors reduces the space of possible choices for the search algorithm~\cite{platt2012}, thus facilitating the study of more fundamental questions.
In fact, despite recent progress, the problem of optimally locating a source within a given time with a prescribed accuracy in a turbulent flow is far from solved~\cite{hutchinson2017_review,reddy2022_review}.

A powerful tool is using the adjoint source-receptor relationship to infer the source location~\cite{pudykiewicz1998,keats2007,wang2019}. However, this technique requires exact knowledge of the underlying equations that govern turbulent odor transport and the possibility to solve them both forward and backward in iterative time, which can be unavailable or impractical in most real-world applications.
On the other hand, for all model-based tools, a key challenge is the practical impossibility of specifying an exact model for the statistics governing odor transport in the presence of turbulence.\\ 
This is a crucial point when using Bayesian approaches that heavily depend on a model of the environment to integrate sensor observations into a probability distribution describing our information about the source. In all real-world applications, we have only partial knowledge of the statistical properties of the environment, which may be arbitrarily complex. Moreover, the inherent complexity of turbulent transport prevents a complete analytical description, even in an idealized physical scenario, and we are inevitably forced to work with a \emph{wrong} model of the environment.
Typical models used in odor search~\cite{ristic2016,hutchinson2018,vergassola2007,masson2009,loisy2022,zhao2022,bosanquet1936,sutton1932} only account for the average transport properties of turbulence in terms of effective advection-diffusion equations. These models are wrong not only because the parameter choices are made arbitrarily but also because the model itself is misspecified.

All in all, these challenges fundamentally limit the utility of models based on Bayesian inference, which are guaranteed to converge to the ground truth only when the model is correctly specified \cite{berk1966,berk1970}. To partially overcome these intrinsic limitations, we explore a new Bayesian approach designed to mitigate the effects of errors in the environmental model. Building upon \emph{many wrongs principle}~\cite{simons2004,berdhal2018}, we introduce a quantity that allows us to rank the models of the environment in the Bayesian framework, providing a principled way to blend information from several (inevitably wrong) models. We show that merging the information gathered from a number of different models --- a procedure that we dubbed \textit{weighted Bayesian update} (WBU) --- helps in reliably inferring the source location. 

In order to test these ideas and the proposed methodology in realistic conditions, we devise a set of direct numerical simulations that mimic the emission of a localized odor source in a three-dimensional turbulent environment. Taking advantage of the Galilean invariance of the Navier-Stokes equation, we model odor transport by Lagrangian tracers advected by the flow with a mean wind, obtaining odor plumes that well approximate those observed within a thin layer at a fixed height in the atmosphere~\cite{crimaldi2001,celani2014}. Here, for the sake of a first benchmark of the localization algorithms, we shall assume that the source lies on the same plane as the sensors, which can thus detect odor particles contained in a thin slab parallel to the wind [see Fig.\ref{fig:1}(a)]. Then, we use a simplified (effective) model for the environment that depends only on a few parameters and neglects temporal and spatial correlations of the odor signal. In spite of the obvious shortcomings of this model, we show how, by suitably weighting different models (i.e., the same one with different parameters), WBU outperforms widely used Monte Carlo methods~\cite{keats2007,doucet2000} when locating the odor source with a specified level of accuracy. By performing several tests with synthetic data and empirically based (\emph{a posteriori}) models, we trace the origin of the problems of Monte Carlo methods to the effects of correlations in the realistic odor signal, to which the WBU shows superior resilience.

The paper is organized as follows. In Sect.~\ref{sec:methods}, we illustrate the setup where the numerical simulations have been carried out and describe the simplified model of the environment as well as the algorithms employed, namely the weighted Bayesian update and another approach based on Monte Carlo techniques. We then present the results in Sect.~\ref{sec:results}, which is divided into two parts. In the first, Sec.~\ref{sec:realistic}, we discuss and compare the performance of the algorithms in the setting closest to reality. In Sect.~\ref{sec:correlations}, we then systematically study how model inaccuracy affects the quality of the source localization. Finally, we summarize our findings and provide an outlook for future studies in Sect.~\ref{sec:discussion}.

%However, the performances of existing algorithms have been benchmarked using simplified models for the turbulent environment far from realistic scenarios.
%how dealing with a wrong model of the environment affects the performance of state-of-the-art algorithms for source localization. 
%In this regard, we shall test their potential applicability by identifying a robust stopping criterion in the search.

\begin{figure}[t!]
\centering
\includegraphics[width=0.95\textwidth]{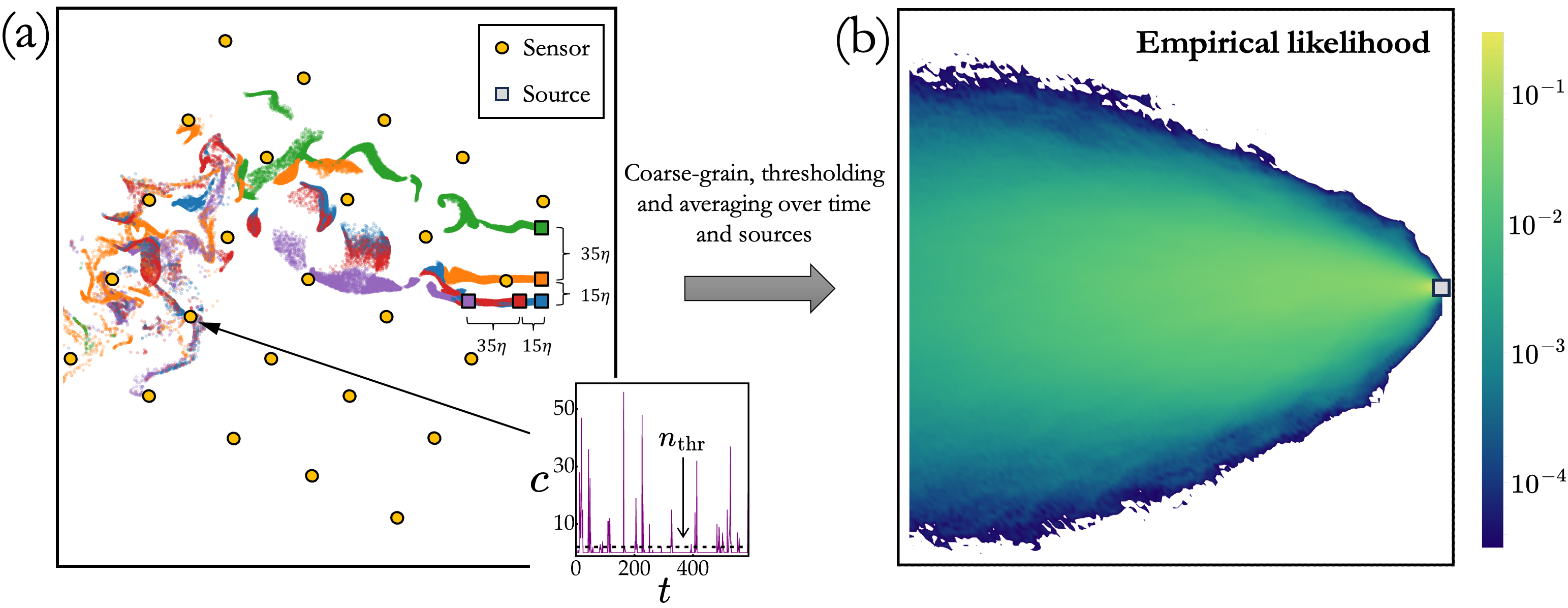}
\caption{\justifying (a) A snapshot of direct numerical simulation (DNS) of the incompressible, 3-D Navier-Stokes equations showing relative positions of sources (squares) and particles color-coded according to the source from which each was emitted. Here, we show particles in a thin slab parallel to the wind and containing the sources, and we have zoomed in close to the sources. Their spatial separation is indicated in units of the Kolmogorov scale $\eta$ of the flow. We have also inserted a network of static sensors (yellow circles) in the same plane, whose measurements will be used in the following to infer the source location, as well as an exemplary time series of the concentration of particles emitted by the purple source and detected by a sensor in the position indicated by the arrow. Please refer to the supplementary video \emph{movie$\_$dns.mp4} (caption in Appendix~\ref{app:supp_movies}) for a 3--D animation of the evolving odor plume obtained from the DNS.
(b) Probability map of making a detection. Hereafter referred to as \emph{empirical likelihood}, such distribution is the result of coarse-graining on a square lattice all the odors' trajectories obtained from the DNS of the 3--D Navier-Stokes equations, then setting a threshold $n_{\rm thr}=2$ on the number of detectable particles, and finally averaging over time the signal of all the five sources (properly placed, by means of a suitable shift, in the same position as indicated by the grey square). Its relevance in our analysis will become clearer in Sec. \ref{sec:correlations}.}
\label{fig:1}
\end{figure}

\section{Methods} \label{sec:methods}

\subsection{Setup of the numerical simulations} \label{subsec:setup}

Let us assume that our objective is to locate the position $\bm{r}_{\rm s}$ of an odor source that emits at a rate $Q$ within a two-dimensional square grid $\Omega$ of size $L\times L$ and lattice spacing $\Delta x$. The odor particles are swept away by the underlying turbulent flow characterized by a mean wind in a given direction $\hat{\bm{U}}$. We then placed a number $N_{\rm s}$ of static sensors in this environment. 
We arrange the sensors in a grid that, assuming no prior knowledge of the wind direction, will be typically rotated by an angle $\theta$ with respect to the latter, as depicted in~Fig.~\ref{fig:1}(a).

%old version, less detail
%The trajectories of the odor particles were produced by state-of-the-art direct numerical simulation (DNS) of the incompressible, 3--D Navier-Stokes equations with $Re_\lambda\simeq 150,$ using a pseudospectral method on a $1024 \times 512\times 512$ grid, with periodic boundary conditions. The system was randomly, isotropically forced at large scales and advected by a uniform mean flow $U=3$ (for reference, $\tilde u_{\rm rms} \simeq 1.22$ in all simulations). Five point sources (see Fig.~\ref{fig:1}) each emitted 1000 Lagrangian tracer particles in a small radius every ten simulation timesteps, or about every $1/15 \tau_\eta$, where $\tau_\eta$ is the Kolmogorov timescale. The particles lived in an infinite domain comprising a lattice of copies of the flow, and their positions were tracked and dumped every $\tau_\eta.$ 

Instead of using an odor concentration field, we model it in terms of particles advected by a turbulent flow. Each particle can be thought of as a patch of odor, or one can consider the number of particles in a given small region as an estimate of the odor concentration.
We produced realistic trajectories of odor particles using state-of-the-art direct numerical simulation (DNS) of the incompressible, 3--D Navier-Stokes equations
\begin{align}
\partial_t \bm{u} + (\bm{u} \cdot \nabla) \bm{u} &= - \nabla p + \nu \nabla^2 \bm{u} + f, \\
\nabla \cdot \bm{u} &= 0,
\end{align}
under turbulent conditions with $\mathrm{Re}_\lambda \simeq 150.$ Here, $f$ is a random, Sawford-type \cite{sawford1991} isotropic forcing at the smallest nonzero wavenumbers of the system, with a correlation time of 160 simulation timesteps. Using a pseudo-spectral code dealiased according to the two-thirds rule, the system was solved on a $1024\times512\times512$ grid, with a uniform spacing $\delta x = \delta y = \delta z \simeq \eta$ (with $\eta$ the Kolmogorov scale), and periodic boundary conditions in all three directions (see also Table~\ref{tab:dnsE} in Appendix~\ref{app:num_sims} for the relevant parameters). The time-stepping was performed using the second-order explicit Adams-Bashforth method. The system was advected by a uniform mean wind $\bm{U} \approx -2.5 u_{\rm rms} \hat{x},$ where $u_{\rm rms}$ is the rms speed of the flow in the comoving frame of the wind, and $\hat{x}$ is the elongated axis of the grid. We produced the mean wind by means of a Galilean transformation. The turbulent intensity achieved by the DNS is strong enough to make the problem highly complex concerning the properties of the underlying flow (refer to Fig.~\ref{fig:dnsE} in Appendix~\ref{fig:appB} for a plot of the spectrum and energy flux). However, this is not meant to reproduce any realistic turbulent realization in the field, such as the atmospheric boundary layer, that would require a Large Eddy Simulation approach and a completely different setup~\cite{bose2018}. 

\begin{figure}[t!]
\centering
\includegraphics[width=\textwidth]{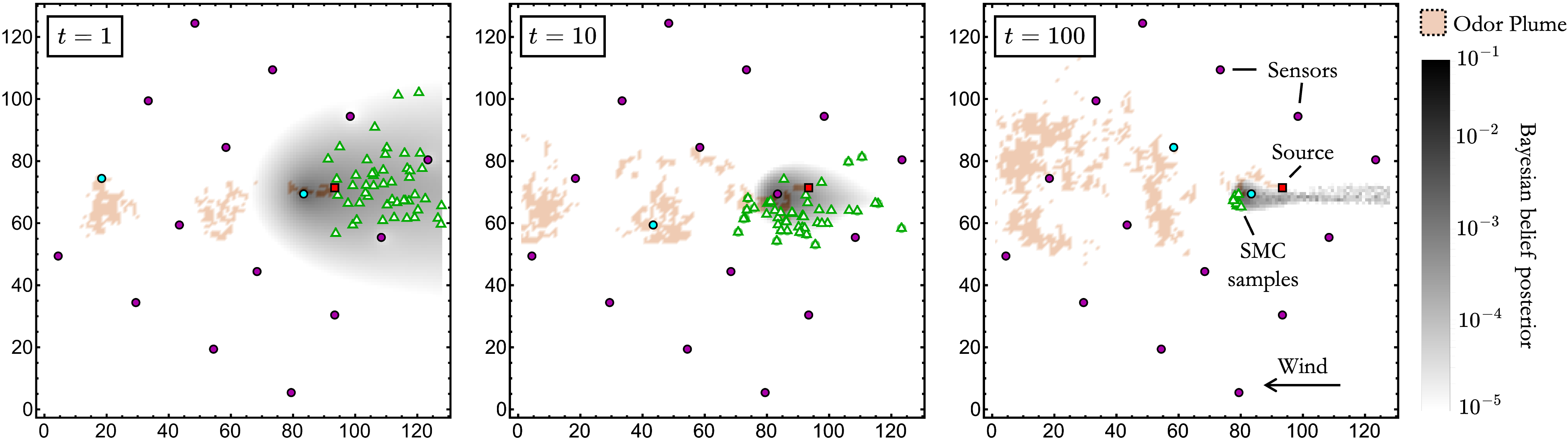}
\caption{\justifying Three time snapshots showing a qualitative comparison between the performance of the two algorithms discussed in this work.
Both approaches use the measurements made by an array of static sensors (circles) looking for a source of odors (red square) advected by a turbulent flow featuring a horizontal wind blowing from right to left. Pink patches indicate the odor plume; greyscale codes the probability of the odor source location obtained from our proposed algorithm based on Bayesian inference (introduced in Sec.~\ref{sec:WBU}); green open triangles depict the candidate source positions yielded by a sequential Monte Carlo (SMC) sampling method. Time here is in units of observations made by each sensor, which is hereafter assumed to be equal to the Kolmogorov timescale $\tau_\eta$ of the flow.}
\label{fig:2}
\end{figure}

The odor particles were modeled as Lagrangian tracer particles, which were emitted by five stationary point sources~[Fig.~\ref{fig:1}(a)]. Each source emits 1000 particles every 10 simulation timesteps, which corresponds to every $\approx 1/15$ Kolmogorov times $\tau_\eta$. The particles evolve within the inertial range of scales, which causes complex spatiotemporal correlations with sparse and intermittent odor distributions [see inset in Fig.~\ref{fig:1}(a)]. We refer the reader to Ref.~\cite{celani2014} for an in-depth discussion of odor landscapes in turbulence.

The fluid velocities $\bm{u}$ at the positions of the particles were obtained using a sixth-order B-spline interpolation scheme and then used to evolve the particle positions $\bm{X}$ in time according to $\dot{\bm{X}}=\bm{u}(\bm{X},t)$ over an infinite lattice of copies of the periodic flow. Their positions, velocities, and accelerations were tracked and dumped every $\tau_\eta$, for a total of 3015 time steps. Each source of particles was treated as independent and we averaged our results over them to achieve better statistics.

To simulate realistic environmental conditions, we then coarse-grain the concentration of particles within a thin slab containing the source. In detail, the slab has thickness $\Delta x$ and transverse extension $128\Delta x\times128\Delta x$, with $\Delta x \approx 15 \eta$. Then, we set a threshold $n_{\rm thr} = 2$ on the number of particles above which sensors make a detection, such as to have a reasonable rate of encounters. Several snapshots of the resulting odor dispersion in the aforementioned arena $\Omega$ are shown in Fig.~\ref{fig:2}, where the pink patches indicate the odor plume. On the one hand, we have explored different choices of threshold and slab thickness, which yielded qualitatively similar results to those presented here. On the other hand, a systematic investigation of larger thresholds or thicker slabs is out of the scope of this work, although potentially interesting.

%Any stochastic model is unable to capture all the nontrivial statistical (temporal and spatial correlations) and topological properties of concentration fields transported by complex flows~\cite{crimaldi2001,celani2014}, making the problem of inferring the source location hard.

\subsection{Model of the environment}\label{subsec:model}

To localize the source within this setting given the sensors' measurements, we shall now introduce a model that captures the mean-flow and mean-diffusion properties of the environment.
If, on the one hand, we assume that we have statistical knowledge of the environment through, for example, the history of previous measurements in the field, we can empirically compute what the probability of detecting an odor particle from the time average of the DNS data is [Fig.~\ref{fig:1}(b)]. We will hereafter refer to this distribution as the \emph{empirical likelihood}, in accordance with Bayesian nomenclature. 
The use of such likelihood has the advantage of avoiding the complication of fitting some environmental parameters while looking for the source. However, it relies on far more detailed prior knowledge of the environment, which is not generally available. On the other hand, we can still take advantage of it to understand the strengths and weaknesses of the localization algorithms, as discussed in Sec.~\ref{sec:correlations}.

In a realistic setup, we are therefore compelled to use a statistical description of odor encounters in a turbulent flow to infer the source location.
Assuming to only know the mean wind direction and have a rough estimate of the turbulence intensity without knowing the details of the turbulent velocity field, we model the turbulent transport of odor particles emitted at rate $Q$
by a point source as an effective advection-diffusion
process~\cite{vergassola2007,hutchinson2018,loisy2022}
\begin{equation}
    \partial_t c + \bm{U}\cdot\boldsymbol{\nabla}c = D\nabla^2c + Q\delta(\bm{r}-\bm{r}_{\rm s}) - c/\tau \, ,
    \label{eq:conc}
\end{equation}
where $c$ is the odor concentration field, $\bm{U}$ the mean wind featured by the turbulent flow, $\tau$ the lifetime of the odors, and $\bm{r}_{\rm s}$ the source position. Note that the combination of molecular and turbulent diffusivity (due to the flow velocity fluctuations) is here described by a single effective (eddy) diffusion coefficient $D$~\cite{biferale1995}. Despite being a strong oversimplification that ignores important multiscale, non-Gaussian properties of the underlying turbulence, as well as spatiotemporal fluctuations in the scalar advection, this model reasonably captures the mean field properties of the odor concentration~\cite{celani2014}.
In the stationary regime, Eq.~\eqref{eq:conc} has an analytical solution, which in three dimensions reads (see, e.g.,~\cite{vergassola2007})
\begin{equation}
    c(\bm{r}-\bm{r}_{\rm s}) = \frac{Q}{4\pi D \lVert\bm{r}-\bm{r}_{\rm s}\rVert} \exp \left[ \frac{(\bm{r}-\bm{r}_{\rm s})\cdot\hat{\bm{U}}-\lVert\bm{r}-\bm{r}_{\rm s}\rVert}{\lambda} \right] \, ,
    \label{eq:conc_sol}
\end{equation}
where we have assumed $\tau\gg1$. A key advantage of the adopted model is that it essentially depends only on three environmental parameters, i.e., the source emission rate $Q$, the mean wind direction $\hat{\bm{U}}$, and the characteristic length scale of the flow $\lambda\equiv 2D/U$.
To mimic the sparseness and intermittency of the odor signals observed from the DNS, and typical of turbulent environments, we assume that the detection is a random process modeled in terms of a Poissonian process with mean $\mu$~\cite{vergassola2007,ristic2016}.
Therefore, the probability that a sensor makes a detection is given by
\begin{equation}
    p(h_i|\bm{r}_i-\bm{r}_{\rm s}) = \frac{[\mu(\bm{r}_i-\bm{r}_{\rm s})]^{h_i}\exp[-\mu(\bm{r}_i-\bm{r}_{\rm s})]}{h_i!}  \, ,
    \label{eq:poisson}
\end{equation}
where $h_i$ is the number of odor particles detected by the $i-$th sensor.
The mean number $\mu$ of particles hitting, within a time interval $\Delta t$,  the $i$-th sensor is related to the mean concentration (\ref{eq:conc_sol}) via the classical Smoluchowski formula~\cite{smoluchowski1918versuch}
\begin{equation}
    \mu(\bm{r}_i-\bm{r}_{\rm s}) = 4\pi a D \Delta t \, c(\bm{r}_i-\bm{r}_{\rm s}) = \frac{Q}{d_i} a\Delta t \exp \left[ \frac{(\bm{r}_i-\bm{r}_{\rm s})\cdot\hat{\bm{U}}-d_i}{\lambda} \right] \, ,
    \label{eq:mu}
\end{equation}
where $\bm{r}_i$ stands for the $i-$th sensor position, and  $d_i \equiv \lVert\bm{r}_i-\bm{r}_{\rm s}\rVert$, $a$ is the sensors' radius.
Hereafter, we will actually assume that each sensor can detect the presence of odors only if the number of particles within its radius $a=\Delta x/2$ exceeds a particular threshold. In other words, the sensors can only perform binary measurements (i.e., $h_i = \{0,1\}$), such that the probability of detection~\eqref{eq:poisson} simplifies into 
\begin{equation}
    \begin{cases}
        p(0|\bm{r}_i-\bm{r}_{\rm s}) = \exp[-\mu(\bm{r}_i-\bm{r}_{\rm s})] \\
        p(1|\bm{r}_i-\bm{r}_{\rm s}) = 1 - \exp[-\mu(\bm{r}_i-\bm{r}_{\rm s})]\, .
    \end{cases}
    \label{eq:poisson_binary}
\end{equation}
We show in Fig.~\ref{fig:3new} an example of a time snapshot of the odor particles generated by such a model of the environment [panel (a)] as well as the resulting detection probability map [panel (b)]. Although it does not grasp the spatiotemporal correlations of the odor signal due to the simplifying assumptions behind it, the model just described can roughly capture the mean field properties of the empirical map obtained from the odors' trajectories resulting from the DNS of the Navier-Stokes equations [see Fig.~\ref{fig:1}(b)]. In the following, we shall use this as the only functional shape for the model of the environment and define \emph{many wrong} models by varying its two free parameters, i.e., the emission rate $Q$ and the characteristic length scale of odors' dispersion $\lambda$, while assuming to know the mean wind direction $\hat{\bm{U}}$. 
It is worth remarking that there is no combination for the value of $Q$ and $\lambda$ which can fit well the empirical distribution [Fig.~\ref{fig:1}(b)] over the whole space. Indeed, values that reasonably capture the probability of making a detection close to the source perform badly in the downwind region and vice versa. This already suggests that a good solution would be to blend the results obtained by using a number of these environmental parameter combinations. In the next section, we discuss a principled way to perform such a blending by showing how we can rank the models of the environment in the Bayesian framework.

\begin{figure}[t!]
\centering
\includegraphics[width=0.7\textwidth]{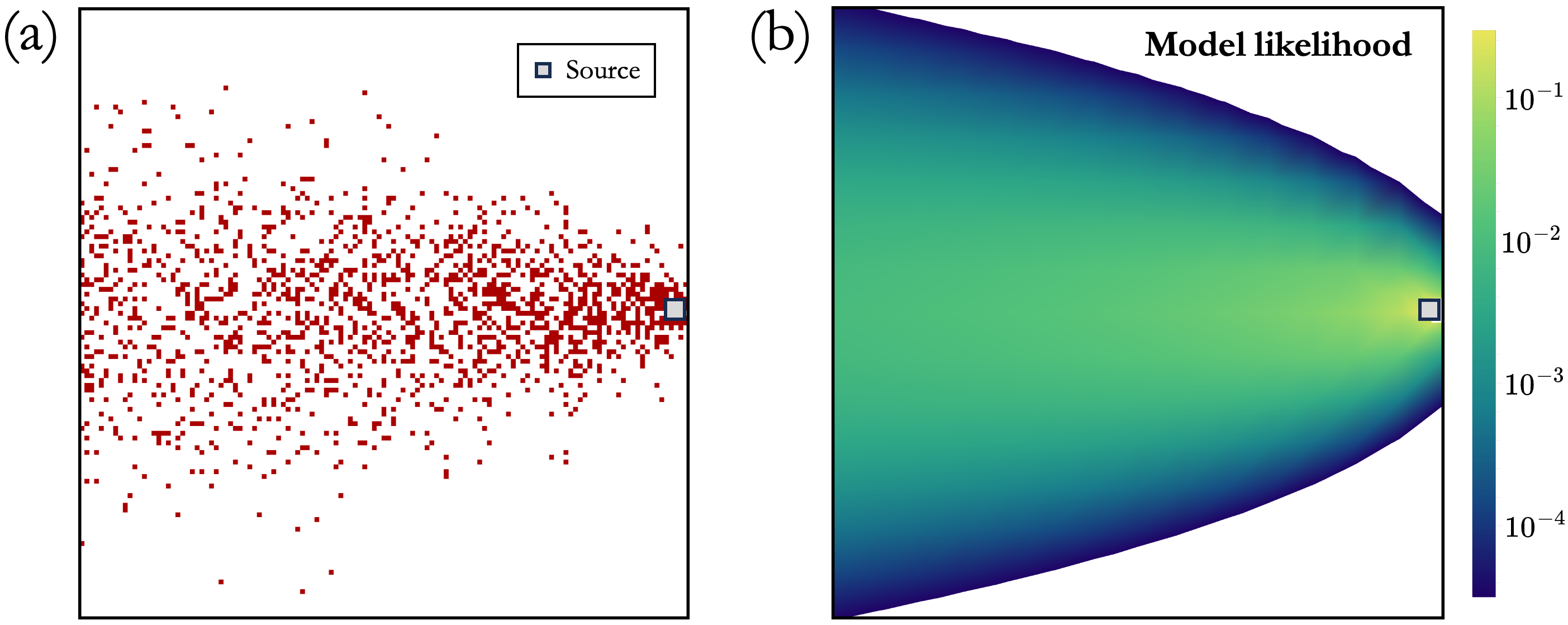}
\caption{\justifying (a) Time snapshot of the odor particles (red dots) emitted by a stochastic source (grey square). (b) Probability map of making a detection according to the stochastic model defined in Eqs.~\eqref{eq:poisson}-\eqref{eq:poisson_binary} for the following choice of parameters: $Q=3$, $\lambda=2.7$, and $\hat{\bm{U}} = -\hat{\bm{x}}$.}
\label{fig:3new}
\end{figure}

\subsection{Weighted Bayesian update: a new way to rank and exploit \emph{many wrong} models \label{sec:WBU}}

Thus, each measurement made by each sensor provides information about the position of the source $\bm{r}_{\rm s}$, which can be processed by employing Bayesian inference. The whole set of measurements from all sensors can be used to update a probability map -- the ``posterior'' or ``belief'' in Bayesian jargon -- of the source location defined over the whole arena $\Omega$, i.e. $B(\bm{r})\equiv \mathrm{Prob}(\bm{r}_{\rm s}=\bm{r})$, which we shall dub as \emph{common belief} to emphasize that it exploits all sensor detections. Since we assume no prior knowledge other that the source is for sure within the arena $\Omega$ (as implicitly implied by the fact that the belief is normalized to unity) and that the source cannot be at the same location as one of the sensors, the belief is always initialized to a uniform distribution and set to zero only in the sensors' positions.
Assuming the simultaneous measurements made by all the $N_{\rm s}$ sensors at time $t$ are independent, the overall conditional probability of a set of observations $\bm{h}^{(t)}$ for a possible given source position $\bm{r}$ is simply a product:
\begin{equation}
    \mathcal{L}(\bm{h}^{(t)}|\bm{r}) = \prod\limits_{i=1}^{N_{\rm s}}p(h_i^{(t)}|\bm{r}_i-\bm{r}) \, ,
    \label{eq:likelihood}
\end{equation}
which depends on the model of the environment as specified in Eq.~\eqref{eq:poisson_binary}, and is also known as the \emph{likelihood} function in Bayesian terminology.
Given a sequential process like the one at hand, at every time step (i.e., once all sensors have performed a measurement), the belief is updated following Bayes' rule~\cite{box2011}
\begin{equation}
    B^{(t)}(\bm{r}) = \frac{\mathcal{L}(\bm{h}^{(t)}|\bm{r}) B^{(t-1)}(\bm{r})}{\int_\Omega {\rm d}\bm{r'} \, \mathcal{L}(\bm{h}^{(t)}|\bm{r'}) B^{(t-1)}(\bm{r'})} \, ,
    \label{eq:bayes1}
\end{equation}
where time $t$ is hereafter measured in units of observations made by each sensor.

Under this update rule, the belief is guaranteed to converge to the correct solution as long as the correct environment model is deployed~\cite{schwartz1965}.
However, this is a rather uncommon case in any realistic scenario, and the model used to update the common belief will always be wrong.  Furthermore, wrong models are indistinguishable from one another during the search since they always make the belief converge to a source position, regardless of whether that is the right one or not.
Therefore, finding a way to rank the models and assess their reliability is of great relevance for any application.

In order to introduce the quantity we used for ranking the models, it is useful to rewrite Bayes' update rule~\eqref{eq:bayes1} in a different way:
\begin{equation}
    B^{(t)}(\bm{r}) = \frac{\prod\limits_{i=1}^{N_{\rm s}} b_i^{(t)}(\bm{r})}{\mathcal{Z}^{(t)}} \, , \;\;\;\;\;\; b_i^{(t)}(\bm{r}) = \frac{p(h_i^{(t)}|\bm{r}_i-\bm{r})  b_i^{(t-1)}(\bm{r})}{\int_\Omega {\rm d}\bm{r'} \, p(h_i^{(t)}|\bm{r}_i-\bm{r'}) b_i^{(t-1)}(\bm{r'})} \, ,
    \label{eq:bayes2}
\end{equation}
where $$\mathcal{Z}^{(t)}\equiv\int_\Omega {\rm d}\bm{r'} \, \prod\limits_{i=1}^{N_{\rm s}} b_i^{(t)}(\bm{r'}) \, .$$
That is to say, owing to the assumption that all the sensors make independent measurements and start with the same prior~\footnote{Actually, as clear from the definition of the common belief $B$ in Eq.~\eqref{eq:bayes2}, for the equivalence between Eq.~\eqref{eq:bayes1} and Eq.~\eqref{eq:bayes2} to hold, it is enough that at time $t=0$ the common prior, i.e. $B^{(0)}$, is equal to the product of the \emph{private} priors $\prod\limits_{i=1}^{N_{\rm s}}b_i^{(0)}$.}
%, owing to the assumption of independence between measurements, 
The common belief $B$ can be built from the superposition of the \emph{private} beliefs (i.e. the belief that can be constructed for each sensor only on the basis of its own measurements history) $b_i$ of the $i=1,...,N_{\rm s}$ sensors, each of which is updated independently at every time step via Bayes' rule. There is a key advantage in rewriting Eq.~\eqref{eq:bayes1} into \eqref{eq:bayes2}, which lies in the interpretation of the update formula.
%The key advantage of this change of perspective lies in the interpretation of the update formula [Eqs.~(\ref{eq:bayes1}- \ref{eq:bayes2})].
Indeed, the hitherto overlooked normalization constant $\mathcal{Z}$ essentially quantifies how much the sensors' \emph{private} beliefs $b_i$  agree on a source position given their measurements and the model at hand. In fact, in Appendix~\ref{app:analytical}, we show analytically that this quantity achieves an asymptotic global maximum when the model is exact (even in the presence of correlations between observations). We also show numerically that this quantity is typically smaller the farther the model is from the ground truth.

This makes $\mathcal{Z}$, hereafter referred to as the \emph{overlap integral}, an ideal candidate to achieve our task as we could use it to weigh different models of the environment. To this end, let us scan the parameter space of the stochastic model introduced above [see Eqs.~\eqref{eq:mu}-\eqref{eq:poisson_binary}] assuming to know only the mean wind direction $\hat{\bm{U}}$, which can be always measured in practice using an anemometer, and run the Bayes update~\eqref{eq:bayes1} independently for each set of parameters $\bm{P}\equiv\{Q,\lambda\}$. Then, we shall blend the information collected into a master belief as
\begin{equation}
     \mathcal{B}^{(t)}_{\rm M}(\bm{r}) = \frac{\prod\limits_{j=1}^{M} \left[B_j^{(t)}(\bm{r}) \right]^{\beta_j^{(t)}}}{\int_{\Omega} {\rm d}\bm{r'} \prod\limits_{j=1}^{M} \left[B_j^{(t)}(\bm{r'}) \right]^{\beta_j^{(t)}}} \, , \;\;\; \beta_j^{(t)} \equiv \frac{\mathcal{Z}_j^{(t)}}{\sum\limits_{j=1}^{M} \mathcal{Z}_j^{(t)}} \, ,
    \label{eq:wbu}
\end{equation}
where $M$ is the total number of distinct sets $\bm{P}_j=\{Q_j,\lambda_j\}$ of model parameters considered, and the index $j$ refers to which of these sets was used to obtain the common belief $B_j^{(t)}$ at time $t$ and the corresponding overlap integral $\mathcal{Z}_j^{(t)}$. 

Equation~\eqref{eq:wbu} is in the spirit of a class of methods called ``generalized Bayesian inference'' which deal with model misspecification
by exponentiating the likelihood by a learning rate before combining it with the prior --- see~\cite{bissiri2016,wu2023} and references therein. Note that our choice to assimilate beliefs in this way is largely heuristic; another possibility would be to perform a weighted average of beliefs with weights $\beta_j^{(t)}$ (we have tested this alternate approach and found it yields similar results). Also note that our $\beta_j^{(t)}$ are closely related to the Akaike weights, a classical information-theoretical tool used for multimodel inference \cite{burnham2002_book}---see Appendix~\ref{app:analytical}.

The algorithmic procedure described in Eqs.~\eqref{eq:bayes1}$-$\eqref{eq:wbu} thus provides a principled way to define a single master belief $\mathcal{B}_{\rm M}$, i.e., the probability distribution about the source location, which is obtained from a weighted average over the results obtained by different models. We will hereafter refer to this approach as the weighted Bayesian update (WBU), whose numerical implementation is detailed in the Appendix~\ref{app:num_sims}.

\subsection{Sequential Monte Carlo with Importance Sampling}

The above-described WBU method offers a new perspective on the implementation of algorithms for odor source localization in turbulent flows. 
However, there exists already a large literature on the possible methods to address the same kind of problem~\cite{hutchinson2017_review}. One of the most commonly used approaches is that based on the use of Monte Carlo sampling methods to estimate the belief~\cite{keats2007,ristic2016,doucet2000}. It is, therefore, natural to ask how the WBU's performance compares with such more conventional approaches.

Due to the wide range of potential applications~\cite{fishman1996}, much research has been conducted over the last decades to make these algorithms more and more computationally efficient and accurate, and, to date, there exist many different variants depending on the specific setup at hand. We refer the reader to~\cite{hutchinson2017_review} for a comprehensive review of the topic. 

In the following, we shall use a state-of-the-art version of the so-called \emph{sequential Monte Carlo} (SMC) algorithm that also involves a Markov chain Monte Carlo (MCMC) perturbation step~\cite{johannesson2004,elfring2021}. Using such an approach, we can simultaneously infer both the source position $\bm{r}_{\rm s}$ and the (unknown) parameters $\bm{P}$ of the stochastic model of the environment. Let us therefore first define \emph{sample} $\bm{\theta}_i$ as a possible combination of such source parameters, that is, $\bm{\theta}_i\equiv\{\bm{r}_{{\rm s},i},\bm{P}_i\}$. 
At each time step, after all sensors have performed a measurement, a collection of $N$ of such samples is drawn from the current belief defined in the $\bm{\theta}$ space, and we assign a weight $w$ to each of them equal to the likelihood of the latest measurement as defined in Eq.~\eqref{eq:likelihood}.
Upon normalization of the sample weights, we shall then compute the effective sample size $N_{\rm eff}$ to avoid the so-called \emph{degeneracy problem}~\cite{elfring2021}. Indeed, if $N_{\rm eff}$ goes below a given threshold $N_{\rm thr}$ (typically set to $N/2$), then it is necessary to generate a new set of $N$ samples. This is known as the \emph{resampling} step. There, for each sample $\tilde{\bm{\theta}}_i$, a new one is selected from the same pool and with a probability equal to its weight that replaces the former. After resampling, all weights are set equal to $1/N$.

Next, we have the so-called Metropolis-Hastings MCMC perturbation step, which consists of moving each sample in its neighborhood and deciding whether to accept or reject the new proposal based on some acceptance criteria~\cite{metropolis1953}. This is essentially done to diversify the $N$ samples and therefore improve the sampling efficiency of the Monte Carlo algorithm~\cite{elfring2021}. 
More specifically, starting from one of the samples at time $t$, say $\tilde{\bm{\theta}}_i^{(t)}$, a Markov chain of length $K$ is generated, where new inferences $\hat{\bm{\theta}}^{(t)}$ are drawn from the previous link in the chain $\tilde{\bm{\theta}}_{i,j-1}^{(t)}$, using a proposal distribution $q(\hat{\bm{\theta}}^{(t)}|\tilde{\bm{\theta}}_{i,j-1}^{(t)})$. Although there exist several valid choices for this distribution~\cite{musso2001}, here we shall use a Gaussian with a mean $\mu=\tilde{\bm{\theta}}_{i,j-1}^{(t)}$ and variance $\sigma^2$ as a free hyperparameter, i.e. $q(\hat{\bm{\theta}}^{(t)}|\bm{\theta}^{(t)}_{i,j-1}) = \mathcal{N}(\bm{\theta}^{(t)}_{i,j-1},\sigma^2)$.
Once the new sample is generated, we shall compute the so-called \emph{acceptance ratio}~\cite{johannesson2004}
\begin{equation}
    \alpha = \prod\limits_{k=1}^t \left[\frac{\mathcal{L}(\bm{h}^{(k)}|\hat{\bm{\theta}}^{(k)})}{\mathcal{L}(\bm{h}^{(k)}|\bm{\theta}^{(k)}_{i,j-1})}\right] \frac{q(\bm{\theta}^{(t)}_{i,j-1}|\hat{\bm{\theta}}^{(t)})}{q(\hat{\bm{\theta}}^{(t)}|\bm{\theta}^{(t)}_{i,j-1})} \, ,
    \label{eq:accept_ratio}
\end{equation}
which basically amounts to the likelihood's history ratio (also known as the \emph{posterior ratio}) divided by the proposal ratio. Including the latter is necessary to correct a bias in the proposal distribution if it is asymmetric~\cite{johannesson2004}. The proposed sample $\hat{\bm{\theta}}^{(t)}$ will then be accepted as a new link in the chain as long as $\alpha>1$, and with probability $\alpha$ otherwise.

Finally, once all samples have been perturbed, we are left with a new approximation of the belief, which reads
$\tilde{B}^{(t)}(\bm{\theta}) = 1/N \sum_{i=1}^N \delta(\bm{\theta}-\bm{\theta}^{(t)}_i)$.
A detailed step-by-step description of the procedure described above is given in Appendix~\ref{app:num_sims}.

Analogously to what we observed in the Bayesian update, once the SMC algorithm's hyperparameters are properly adjusted, it will systematically converge to the correct source location as long as the model of the environment is functionally exact. 
However, it is not clear how this approach would compare with the WBU one in a realistic scenario where it uses an inevitably misspecified model of the environment.

\begin{figure}[ht!]
\centering
\includegraphics[width=0.75\textwidth]{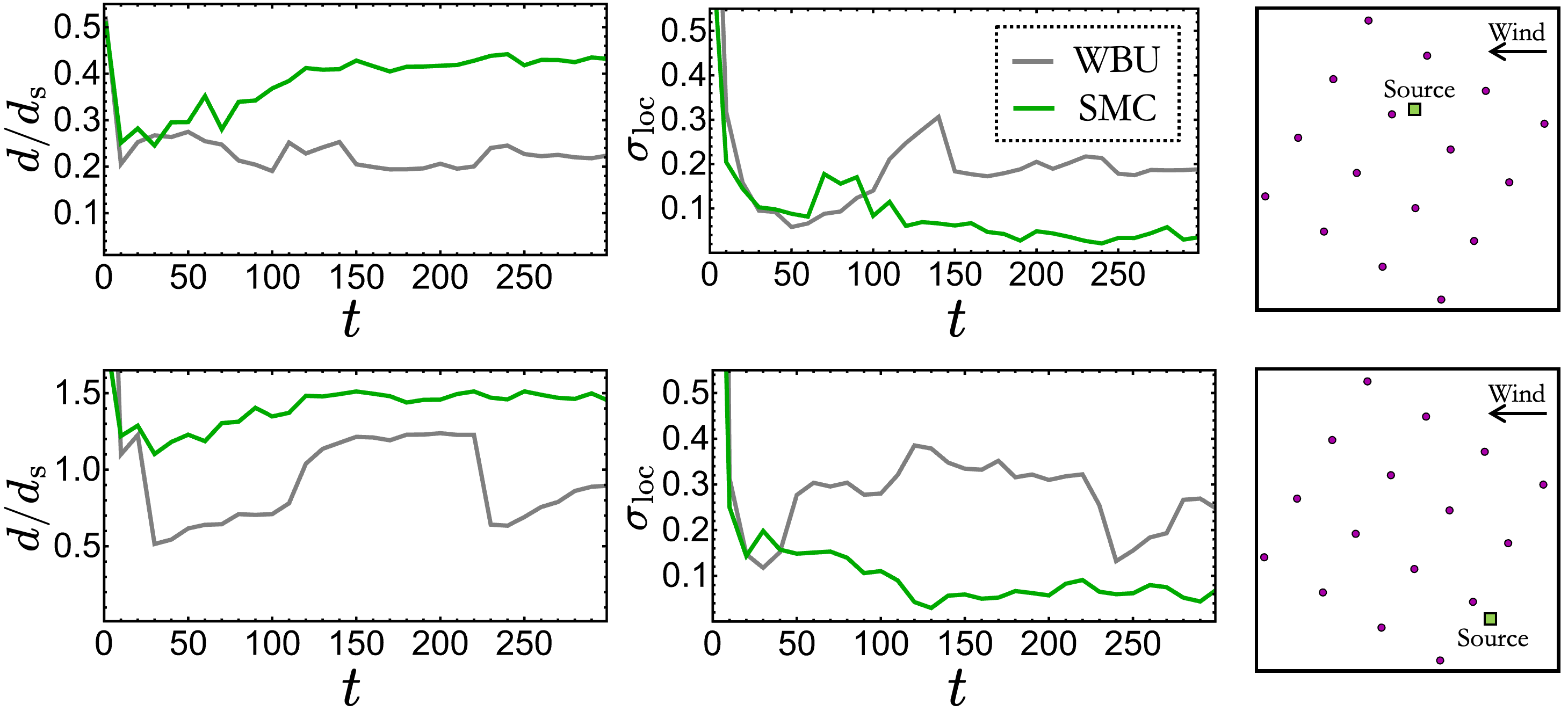}
\caption{\justifying Comparison between weighted Bayesian update (WBU, gray curves) and sequential Monte Carlo (SMC, green curves) in terms of the time evolution of the distance $d$
between the estimated source position and the ground truth and of the corresponding uncertainty $\sigma_{\rm loc}$, both in units of the distance $d_{\rm s}$ between the sensors. The data shown in the two rows correspond to different source-sensor configurations, as indicated in the rightmost panel of each row. Please refer to the supplementary video \emph{movie$\_$algorithms$\_$evolution.mp4} (caption in Appendix~\ref{app:supp_movies}) showing the time evolution of the WBU's master belief together with the SMC's samples during the same episode and configuration shown in the top row.}
\label{fig:4}
\end{figure}

\section{Results} \label{sec:results}

\subsection{Stop criterion for source localization}\label{sec:realistic}

The performance of localization algorithms is typically assessed based on their ability to estimate the position of the odor source with a specified accuracy or within a given time limit~\cite{hutchinson2017_review,thomson2007,ristic2015_accuracy}. 
However, more generally, we shall test the reliability of such algorithms by envisioning their application in a real-world scenario, where we do not know \emph{a priori} where the source is and have to decide when to stop looking for it. 
Therefore, for an algorithm to be effective in practice, it must show a correlation between an observable quantity and the quality of its estimate of the source location. 

\begin{figure}[ht!]
\centering
\includegraphics[width=\textwidth]{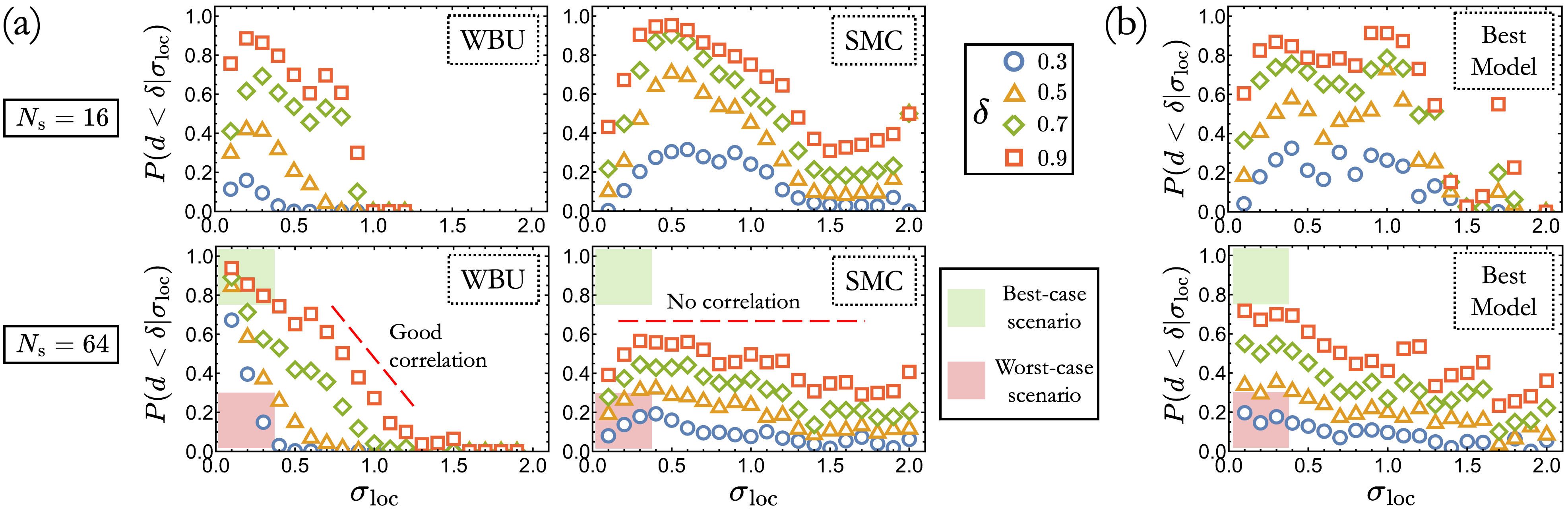}
\caption{\justifying (a) Four panels showing the probability that the distance $d$ between the estimated source position $\bm{\bar{r}}$ and the ground truth $\bm{r}_{\rm s}$ is below a given threshold $\delta$ (indicated in the legend) as a function of the belief/sample standard deviation $\sigma_{\rm loc}$ around the belief/sample mean $\bm{\bar{r}}$. Lengths are in units of the distance between sensors.
The rows differ in the number of sensors, while the two columns correspond to the algorithm used [left: weighted Bayesian update (WBU), right: sequential Monte Carlo (SMC)]. In the two plots at the bottom, we have included a dashed line to visually emphasize the presence (or absence) of correlation between $P$ and $\sigma_{\rm loc}$ when using either algorithm. Moreover, in the same plots, the green and red squares respectively highlight the most and least desirable regions where the curves shown should converge as $\sigma_{\rm loc}\to0$. Indeed, the \emph{best-case scenario} (green square) reflects the situation in which the belief/samples converge to a point very close to the true source position. Conversely, the so-called \emph{worst-case scenario} corresponds to the instance when it ends up with high precision in a wrong location.
The data shown here have been obtained by running both algorithms in a set of $25$ episodes of time length $T=600$ (in units of the Kolmogorov timescale $\tau_\eta$ of the flow) for each of the $30$ different odor source locations considered (details in Appendix~\ref{app:num_sims}).
(b) Same as (a) but computed using the models that ranked first with the overlap criterion discussed in Sec.~\ref{sec:WBU}.}
\label{fig:3}
\end{figure}

To this end, a good candidate quantity is the current uncertainty about the source location. This can be formally defined in the WBU approach as the variance of the master belief \eqref{eq:wbu}, which reads
\begin{equation} \label{eq:sigma_loc}
    \sigma^2_{\rm loc}\equiv\int_\Omega \, {\rm d}\bm{r} \, \mathcal{B}_{\rm M}(\bm{r})(\bm{r}-\bm{\bar{r}})^2 \, ,
\end{equation}
where $\bm{\bar{r}}\equiv\int_\Omega \, {\rm d}\bm{r} \, \mathcal{B}_{\rm M}(\bm{r})\bm{r}$ is the estimated source location. Clearly, the $\sigma^2_{\rm loc}$ and $\bm{\bar{r}}$ counterparts in the SMC algorithm are, respectively, defined as the variance and mean computed over the Monte Carlo samples.

In the ideal case where the model of the environment is exact, $\sigma^2_{\rm loc}$ would be inversely correlated with how close the estimated position is to the ground truth. In other words, the smaller the variance of the belief, the more accurate the source location estimate would be. However, the exact model of the environment is inaccessible in any practical scenario.
Therefore, such a correlation is not guaranteed to hold in general and may depend on the algorithmic procedure employed.
In fact, while the uncertainty about the location of the source (measured by $\sigma_{\rm loc}$) decreases as time passes, regardless of the model used, this is not necessarily true for the distance between the estimated position of the source $\bm{\bar{r}}$ and the actual location. That is to say, a wrong model of the environment would, in general, lead the algorithm to make the belief/samples collapse with high precision in a wrong location.

In Fig.~\ref{fig:5}, we provide a first insight into how the WBU and SMC algorithms perform over time in two exemplary source-sensor configurations, as indicated by the rightmost panels in each row. Specifically, we show the time evolution of their accuracy in estimating the source position (the distance $d$ between $\bm{\bar{r}}$ and the true source position $\bm{r}_{\rm s}$ --- first column) and of the corresponding uncertainty ($\sigma_{\rm loc}$ --- second column). We may observe that in both situations the samples of the SMC algorithm (green curves) tend to collapse over time in a very small region ($\sigma_{\rm loc}\lesssim0.1$), while the distance of the estimated source location from the ground truth quickly saturates to a constant value. This means that SMC has most samples clustered into a wrong source location.
On the other hand, the WBU approach (gray curves) localizes the source with higher accuracy while typically staying rather diffuse ($\sigma_{\rm loc}\gtrsim0.2$). Besides, WBU's uncertainty about the source location shows a much better correlation with the quality of its estimate, contrary to what happens when using the SMC algorithm.

We shall now support this first qualitative picture and quantify the robustness of a given localization algorithm by looking at the probability $P(d<\delta|\sigma_{\rm loc})$ that the distance $d$ between $\bm{\bar{r}}$ and the true source position $\bm{r}_{\rm s}$ is below a given threshold $\delta$, conditioned on the uncertainty on the source estimate as defined in \eqref{eq:sigma_loc}.
In particular, the reliability of a localization algorithm can then be measured by checking whether such probability $P(d<\delta|\sigma_{\rm loc})$ correlates with the only observable quantity, namely the standard deviation $\sigma_{\rm loc}$. 

We find that this is indeed the case for the WBU approach when implemented in the most realistic scenario introduced in Sect.~\ref{subsec:setup}. As shown in Fig.~\ref{fig:3}(a), the probability of locating the source within a distance $d$ smaller than that between sensors is indeed well correlated with the corresponding uncertainty when using WBU (left column). More specifically, it turns out that the smaller the uncertainty $\sigma_{\rm loc}$ about the location of the source, the higher the probability $P$ of being close to the correct position.
However, this does not hold for the more standard SMC algorithm, which hardly shows any correlation between these two quantities (right column). Moreover, this turns out to be independent of the number of sensors deployed ($N_{\rm s}=\{16,64\}$ are shown in the same figure) and of the threshold $\delta$ on the accuracy (different symbols in each plot). 
Actually, we find that when deploying a larger number of sensors (and thus, in principle, when trying to localize the source within a smaller region), the performance of SMC becomes even worse (compare the two top panels with the bottom ones). As can be seen, SMC basically features a flat conditional probability, which means that having a small standard deviation $\sigma_{\rm loc}$ is not correlated with being close to the odor source. In fact, the observed convergence of $P(d<\delta|\sigma_{\rm loc})\to0$ as $\sigma_{\rm loc}\to0$ (see red shaded region) indicates that this algorithm tends to localize the source with great precision in the wrong place. In contrast, the conditional probability of being close to the actual source location as $\sigma_{\rm loc}\to0$ tends to one when using WBU in the same setup (see the green shaded region). This corresponds to the ideal case in which the algorithm converges with very good accuracy at a point close to the source.

It is worth stressing that there is no time conditioning here and the probability $P(d<\delta|\sigma_{\rm loc})$ is obtained by considering all the data available from our numerical simulations regardless of the source configuration. In other terms, at each update and for each instance, we compute the estimated source location $\bar{\bm{r}}$ and the associated uncertainty $\sigma_{\rm loc}$, and then collect all these data exactly to compute the conditional probability shown in these plots. Note that we have analyzed the distribution of $\sigma_{\rm loc}$ values and checked that small localization uncertainties ($\sigma_{\rm loc} < 1$) are sufficiently populated (data not shown), ensuring robust statistical reliability in the results presented here.
We refer the reader to Appendix~\ref{app:num_sims} for further details on the numerical simulations.

Hence, our results suggest that the WBU is a more reliable method than the standard SMC in the sense that, in WBU, the uncertainty about the source location $\sigma_{\rm loc}$ is a more reliable indicator of the proximity of its estimate to the ground truth than it is in SMC.

Furthermore, this has the additional benefit that it allows us to define a robust stop criterion. Indeed, thanks to the observed monotonic decrease of $P(d<\delta|\sigma_{\rm loc})$ vs. $\sigma_{\rm loc}$ when employing WBU, it will be sufficient to look at the value of $\sigma_{\rm loc}$ to know when one has a good chance of finding the source at a sufficiently small distance from the current estimate and stop the search accordingly. In contrast, there is no way to define a robust stop criterion when using the SMC algorithm.

Remarkably, this would no longer hold if we used only the model with the largest overlap integral $\mathcal{Z}$ to infer the source location. Indeed, in such a case, the correlation between the distance from the true source location and its uncertainty almost disappears, with a result comparable to the one of the SMC algorithm (compare Fig.~\ref{fig:3}(b) with the panels of Fig.~\ref{fig:3}(a)). This complies with what was observed at the end of Sec.~\ref{subsec:model}, namely that, given the level of complexity of turbulent odor transport, there is not a single best combination of parameters to model the environment, which further emphasizes the importance of blending information from many wrong models to estimate the source location, as done in the WBU approach.

\subsection{Model misspecification: the effect of spatial and temporal turbulent correlations} \label{sec:correlations}

Although both WBU and SMC use the same (wrong) model of the environment to infer the source location, they have different outcomes, with the former proving to be less sensitive to the model misspecification than the latter. 
Therefore, it is worth investigating more systematically how model errors affect the performance of such source localization algorithms. 
As shown in the table in Fig.~\ref{fig:5}(a), we can consider four possible ways to infer the source location with a given algorithm when dealing with realistic data of turbulent odor dispersion.

The first possibility is to use the Lagrangian time series
obtained from the DNS to determine whether a sensor detects an odor or
not and then deploy a (inevitably misspecified) model of the
environment ---like the one derived from Eq.~\eqref{eq:conc}--- to
interpret such detections and compute the likelihood (\emph{Model
likelihood-Correlated signal} ---MlCs--- top left cell in the table).
This is precisely the scenario we have analyzed so far, which is the
closest one to reality, as it relies on minimal prior knowledge of the
environment and directly uses the Lagrangian data. 

\begin{figure}[t!]
\centering
\includegraphics[width=\textwidth]{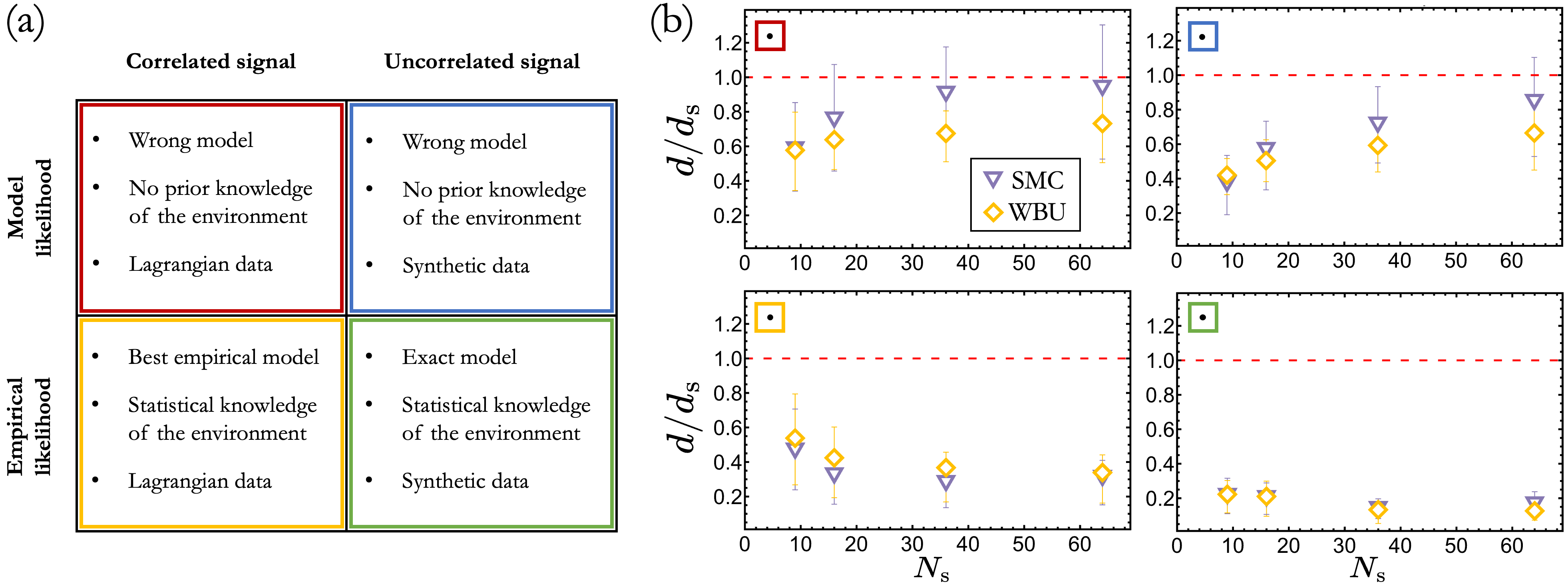}
\caption{\justifying (a) Table summarizing all the possible ways to infer the source location when dealing with realistic data of turbulent odor dispersion. They may differ for the model used to compute the likelihood (rows) or for the signal received by the sensors (columns). (b) Mean distance $d$ from the source at the end of each episode in units of the distance $d_{\rm{s}}$ between sensors as a function of the number of sensors $N_{\rm s}$. Different symbols correspond to the algorithm employed (SMC: inverted purple triangles; WBU: yellow diamonds). Error bars are the first and third quartiles of the distribution. Each plot refers to one of the four cases illustrated in table (a), as indicated by the box color in the upper left corner. Note that the increase in localization error $d$ observed in the top two panels arises from the fact that this quantity is normalized by the distance between sensors $d_{\rm{s}}$. Therefore, as sensor density increases and the inter-sensor distance decreases, the relative error may appear amplified while still being a decreasing function of $N_{\rm s}$ in absolute terms, as one would expect.}
\label{fig:5}
\end{figure}

Alternatively, as a model of the environment, we can use the empirical likelihood [Fig.~\ref{fig:1} (b)] calculated from the time history of all the odor trajectories obtained from the DNS. 
On the one hand, this approach [\emph{Empirical likelihood-Correlated signal} ---ElCs---  bottom left cell in the table in Fig.~\ref{fig:5}(a)] greatly simplifies the search since there are no environmental parameters to fit, and it also represents the best existing (empirical) model one can aim for in practice to infer the source location. 
However, it is still an imperfect model of the environment as it does not capture the time and spatial correlations featured by the odor plume~\cite{celani2014}. 

To study how such correlations affect the source position estimation, we should consider the case where the detections are not directly taken from the DNS time series but instead randomly drawn from the empirical likelihood. Indeed, the sensors' measurements are in this way uncorrelated while still featuring the same detection statistics as the original signal.
At this point, we can decide whether to deploy the empirical likelihood itself to infer the source location, in which case we would be using the exact model of the environment (\emph{Empirical likelihood-Uncorrelated signal} ---ElUs--- bottom right cell in the table) or the usual probabilistic model of detections (\emph{Model likelihood-Uncorrelated signal} ---MlUs--- top right cell in the table). 
The former can be helpful as a benchmark since, provided the number of measurements is large enough, it should show convergence to the correct solution for any properly implemented localization algorithm based on Bayesian inference.
The latter is instead useful compared to the first two cases illustrated above since, in this case, sensors' observations are uncorrelated, and, as a result, it isolates the effect of dealing with a functionally wrong model of the environment. Note that in the ElC or ElU cases (bottom panels of Fig.~\ref{fig:5}(a)), where the model—whether wrong or exact—is fixed, WBU reduces to the standard Bayesian update, as no model blending or ranking occurs. Nevertheless, with some abuse of language, we will still refer to it as WBU in the following.

This completes the picture in the table in Fig.~\ref{fig:5}(a). We are now ready to compare the performance of WBU and SMC in each of the four scenarios just described. To this end, we examine the results reported in Fig.~\ref{fig:5}(b). There, we show the average distance $d$ between the actual source location and its estimate given by either algorithm at the end of distinct episodes of set time length as a function of the number of sensors $N_{\rm s}$ deployed in the search (details in Appendix~\ref{app:num_sims}). 
Some observations are in order. First, both WBU (yellow diamonds) and SMC (purple inverted triangles) show comparable performance within error bars as long as they use the empirical likelihood to model the detection statistics (plots in the bottom row). This is somehow expected as there is no other parameter to infer other than the source position. More specifically, when using the exact model as in the ElUs scenario, both approaches tend to converge to the correct source location as expected (bottom right plot). At the same time, by looking at the ElCs case (bottom left plot), we may observe how much the sole presence of time and spatial correlations in the signal affects the quality of the source estimate. Interestingly, even though the empirical model does not account for such correlations, both algorithms manage to locate the source within a distance $d$ smaller than half of the sensor separation.

Now, let us see what changes when using the probabilistic model introduced in Sect.~\ref{subsec:model} to infer the source location (MlCs and MlUs scenarios ---plots in the top row). Although using the wrong model has obvious shortcomings and further degrades the performance of both approaches, overall,  they prove to be quite robust and still manage, on average, to locate the source within a distance smaller than the one between sensors.
Furthermore, comparing the results obtained with uncorrelated signals (MlUs, top right plot) with the most realistic scenario featuring the correlated signal (MlCs, top left plot), we may notice that the SMC's performance gets substantially worse in the latter case, while WBU is basically unaffected.

\begin{figure}[ht!]
\centering
\includegraphics[width=0.85\textwidth]{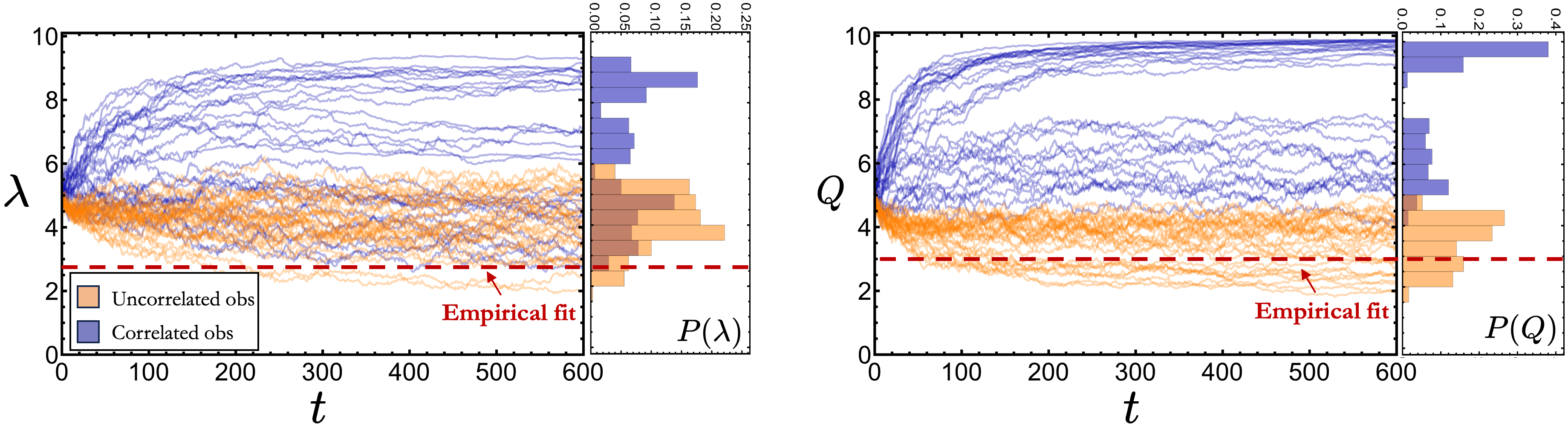}
\caption{\justifying Comparison between the values of the model parameters (left: $\lambda$; right: $Q$) inferred by the SMC algorithm as a function of time when directly using the DNS time series (blue curves, time-correlated signal) and when extracting the detections from the corresponding empirical likelihood (orange curves, uncorrelated signal). In both cases, each curve corresponds to the episodes-averaged value of the parameter obtained in one of the $30$ configurations (different source positions) where the algorithm has been tested. The histograms on the right of each plot have been obtained by considering the inferred values of the parameters at times $t\ge300$, while the horizontal dashed red lines stand for the values of $\lambda$ and $Q$ derived from the best fit of the empirical likelihood. Data shown here correspond to the setup with $N_{\rm s}=16$ sensors.}
\label{fig:6}
\end{figure}

The higher sensitivity of SMC to correlations can be rationalized by looking at the time evolution of the model parameters, namely the emission rate $Q$ and the characteristic length scale of turbulent odor dispersion $\lambda$, inferred by such an algorithm. As shown in Fig.~\ref{fig:6}, when the signal received by the sensors is correlated in space and time (blue curves), SMC has the tendency to overestimate both $\lambda$ and $Q$, even saturating to the maximum allowed value. The picture dramatically changes when the sensors' observations are instead uncorrelated (orange curves). There, the inferred values of the parameters are much closer to the ones obtained from the best fit of the empirical likelihood (dashed red line in both plots), which means the algorithm is in this case pointing in the right direction.

Hence, the observed sensitivity of the SMC algorithm in the most realistic scenario lies in the fact that it must perform a real-time inference of the model parameters (while looking for the source), which is greatly impacted by the presence of time correlations in the signal. However, this is not the case for the WBU approach since it uses a predetermined and discrete set of parameters, then runs the different models independently, and only at the end merges their outcome based on the current ranking provided by the overlap integral of each model. While the WBU algorithm dynamically ranks models during inference, we shall stress that these rankings are highly configuration-dependent and do not reflect intrinsic model validity. Thus, the \emph{static} nature of the parameter space and the use of all available (misspecified) models are thus the defining strengths of the newly introduced approach.

\section{Discussion} \label{sec:discussion}

In this work, we revisit the problem of locating an odor source in a realistic turbulent environment with a network of static sensors. To this end, we used DNS of the incompressible 3--D Navier-Stokes equations to simulate turbulent odor dispersion in the presence of a strong mean wind. We then used the data of the Lagrangian particle trajectories to systematically investigate how the errors made in modeling such an environment, a problem unavoidable in field applications, affect the performance of source localization algorithms based on Bayesian inference.
In particular, we used a class of effective models, often used in odor source localization applications, which assumes minimal knowledge of the transporting velocity field, namely the presence of a mean wind and a rough notion of the turbulent intensity. Such models, which end up depending only on two free parameters, manage to roughly capture the average properties of the odor concentration field despite completely missing all the nontrivial spatiotemporal correlations of realistic odor plumes.

Within this framework, we identified a quantity that effectively ranks different models of the environment just on the basis of the history of observations made by the sensors in the field. 
This is what we call \emph{overlap integral} $\mathcal{Z}$ because it essentially measures the degree of consensus among sensors on a single source location. 
This quantity, which is maximized when the model is exact, is also proportional to the posterior estimate of the probability of making the sequence of observations, integrated over all possible source locations. Thus, our weighted Bayesian update (WBU) approach to source localization may be viewed as a form of maximum likelihood estimation (MLE) applied to the model itself. The use of MLE to quantify the degree of model misspecification was previously studied in an abstract setting in~\cite{white1982}.

%In other words, as this quantity represents the posterior estimate of the probability of making the sequence of observations integrated over all possible source locations, we have been performing a maximum a posteriori estimation applied to the model itself.
%In particular, we were able to analytically show that it features a global maximum when the model at hand is exact, and this, in turn, allowed us to define a new way to infer the source location when the correct model of the environment is unknown: the weighted Bayesian update (WBU).

Through our analysis, we may conclude that WBU is a more robust approach to locating an odor source in a turbulent environment than state-of-the-art methods relying on Monte Carlo sampling techniques.
In fact, our results highlight some fundamental weaknesses of the sequential Monte Carlo (SMC) algorithm, which features a strong sensitivity to the presence of time/space correlations in the sensors' detections, especially in the most general case where it must also infer the model parameters in real-time together with the source location. This is in contrast to WBU, which is insensitive to correlations and proved to be only slightly affected by the use of wrong models. In particular, WBU, as opposed to SMC, is able to maintain the desired correlation between the uncertainty about the source location and the distance of its current estimate from the ground truth. Thus, WBU provides a robust stop criterion for the search.

The main novelty of the WBU approach stems from the idea that merging the information gathered from many possible interpretations of the measurements recorded by the sensors may help compensate for model errors consistent with \emph{many wrongs principle}~\cite{simons2004,berdhal2018}. In the realm of olfactory search, this represents a fundamental stepping stone toward a thorough investigation of the effect of misspecified models on the localization of odor sources in turbulent environments, an unavoidable difficulty in practical applications.

The results and methodology presented have clear potential applications in environmental monitoring~\cite{tariq2021,ullo2020} and early warnings~\cite{esposito2022}, as they allow a reliable identification of a potential area of intervention when some hazardous substances are detected. It would also be interesting to explore the possibility of capitalizing on the use of many wrong models in the case of olfactory search by single~\cite{vergassola2007,loisy2022} or multiple \cite{masson2009} moving agents, where it can help mitigate the model misspecification and thus allow for better decisions of the agent(s). 
Furthermore, given the importance of source localization problems, either with fixed sensors or with moving agents (see e.g.~\cite{reddy2022_review} and references therein), we stress the importance of having a few publicly available databases featuring different turbulent conditions (wind and turbulence intensity, presence or not of inhomogeneities, etc.) to test on the same data different algorithms based on Bayesian approaches, as discussed here, or on some heuristics~\cite{balkovsky2002,durve2020} or machine learning procedures (see, e.g., \cite{rigolli2022learning,verano2023olfactory} for recent attempts using either fixed or moving sensors).

Beyond source localization, the WBU approach is, in principle, also adaptable to other Bayesian inference problems with model misspecification and multiple independent measurements. Although exploring its application in these broader Bayesian contexts is beyond the scope of this paper, it presents an interesting outlook for future research. Finally, we suggest that the overlap integral could be maximized instead by gradient ascent, which would afford a reduction in computational cost and memory load while losing the extra information from running many wrong models. It would be interesting, in particular, to explore the use of a powerful, many-parameter model such as a neural network in such a context.

%Finally, in future investigations, it would be interesting to exploit the information that can be gleaned from the \emph{overlap integral} $\mathcal{Z}$. As shown in our work, this plays the role of a performance index that can be maximized ``online" to derive a more performing model of the environment, possibly deploying, e.g., neural networks or gradient ascent algorithms that would avoid running several models simultaneously.

\section*{Acknowledgments}

We thank C. Calascibetta, N. Cocciaglia, M. Sbragaglia and M. Vergassola for useful discussions. 
We acknowledge financial support under the National Recovery and Resilience Plan (NRRP), Mission 4, Component 2, Investment 1.1, Call for tender No. 104 published on 2.2.2022 by the Italian Ministry of University and Research (MUR), funded by the European Union – NextGenerationEU– Project Title Equations informed and data-driven approaches for collective optimal search in complex flows (CO-SEARCH), Contract 202249Z89M. – CUP B53D23003920006 and E53D23001610006.
This work was  supported by the European Research Council (ERC) under the European Union’s Horizon 2020 research and innovation program (Grant Agreement No. 882340).

\appendix
\setcounter{figure}{0}
\renewcommand\thefigure{A\arabic{figure}}
\setcounter{equation}{0}
\renewcommand\theequation{A\arabic{equation}}
\setcounter{table}{0}
\renewcommand\thetable{A\arabic{table}}

\section{Analytical argument for the convergence of the overlap integral} \label{app:analytical}
Let the true likelihood of the $i$-th sensor be $p^*(h_i|\bm{r}_i - \bm{r})$  and let Bayesian updates be performed with a model $p(h_i|\bm{r}_i-\bm{r})$ (as in Eq.~\eqref{eq:poisson_binary}) not necessarily correctly specified---i.e., $p\ne p^*$ in general. Although most of the analysis will be for arbitrary distributions, we will specialize in the case of binary detections, so that $p$ and $p^*$ are both Bernoulli but with different distributions of the mean detection rate $\mu$. Assume (without significant loss of generality) that the prior is uniform, and let the true source location be $\bm{r}_{\rm s}.$

Let us rewrite the overlap integral by expanding Eq.~\eqref{eq:bayes2}. After $T$ timesteps, we have
\begin{equation}
\mathcal{Z}^{(T)} = \frac{{\cal P}_T}{{\cal N}_T},
\end{equation}
where
\begin{equation}
{\cal P}_T \equiv \int_\Omega d\bm{r} \,  \prod_{i=1}^{N_s} \prod_{t=1}^T p(h_i^{(t)} | \bm{r}_i - \bm{r}) = \int_\Omega d\bm{r} \, \exp \left(\sum_{i=1}^{N_s} \sum_{t=1}^T \log p(h_i^{(t)} | \bm{r}_i - \bm{r}) \right)   
\end{equation}
and
\begin{equation}
    {\cal N}_T \equiv \prod_{i=1}^{N_s} \int_\Omega d\bm{r'} \, \prod_{t=1}^T p(h_i^{(t)} | \bm{r}_i - \bm{r'})  = \prod_{i=1}^{N_s} \int_\Omega d\bm{r'} \, \exp \left( \sum_{t=1}^T \log p(h_i^{(t)} | \bm{r}_i - \bm{r'}) \right).
\end{equation}
 ${\cal P}_T$ may be understood as the posterior likelihood of the sequence of measurements, aggregated from all the sensors. ${\cal N}_T$ is a factor that is inherited from the normalizations of the private sensor beliefs.

We now follow a standard line of reasoning to study the asymptotic forms of $\mathcal{P}_T$ and $\mathcal{N}_T$; see, for example,~\cite{berk1970} for a similar argument. We apply the (weak) law of large numbers (LLN), which implies that the arguments of the exponentials converge in probability to their expectations over observations $h$:
\begin{align}
\mathcal{P}_T \to \bar{\mathcal{P}}_T & \equiv \int_\Omega d\bm{r} \, \exp \left(T\sum_{i=1}^{N_s} \sum_{h_i\in\{0,1\}} p^*( h_i | \bm{r}_i - \bm{r}_{\rm s}) \log p(h_i | \bm{r}_i - \bm{r}) \right) \nonumber \\ & = C_T \int_\Omega d\bm{r} \, \exp \left[-T\sum_{i=1}^{N_s} D_{\rm KL}(p^*( \cdot | \bm{r}_i - \bm{r}_{\rm s}) \parallel p(\cdot  | \bm{r}_i - \bm{r})) \right], \label{eq:pt} \\
\mathcal{N}_T \to \bar{\mathcal{N}}_T & \equiv \prod_{i=1}^{N_s} \int_\Omega d\bm{r'} \, \exp \left(T \sum_{h_i\in\{0,1\}} p^*( h_i| \bm{r}_i - \bm{r}_{\rm s}) \log p(h_i | \bm{r}_i - \bm{r'}) \right) \nonumber \\ &= C_T \prod_{i=1}^{N_s} \int_\Omega d\bm{r'} \, \exp \left[-T D_{\rm KL}(p^*( \cdot | \bm{r}_i - \bm{r}_{\rm s}) \parallel p(\cdot  | \bm{r}_i - \bm{r'})) \right], \label{eq:nt} 
\end{align}
where
\begin{equation}
    C_T = \exp\left(-T \sum_{i=1}^{N_s} H(p^*( \cdot | \bm{r}_i - \bm{r}_{\rm s}))\right),
\end{equation}
$D_{\rm KL}(f(\cdot) \parallel g(\cdot))\equiv \sum_{h=\in \{0,1\}} f(h) \log f(h)/g(h)$ indicates the Kullback-Leibler divergence between distributions $f$ and $g$, $H(f(\cdot))\equiv - \sum_{h=\in \{0,1\}} f(h)\log f(h)$ is the Shannon entropy.
%and we have made use of the identity
%
%\[
%-\sum_i f_i \log g_i = \sum_i f_i \log f_i/g_i - \sum_i %f_i \log f_i = D_{\rm KL}(f\parallel g) + H(f).
%\]
Note that the $C_T$'s in $\mathcal{Z}^{(T)}$ will cancel, which is numerically beneficial.

The application of LLN is valid as long as each observation $h_i^{(t)}$ has a bounded variance and the correlations between observations decay to zero over time (that is, $\mathrm{Cov}(h^{(t)},h^{(t+\tau)}) \xrightarrow{\tau \to \infty}0$)~\cite{bernshtein1918}, so Eqs.~(\ref{eq:pt})--(\ref{eq:nt}) apply even in the presence of spatiotemporal correlations. 

First, we inspect $\bar{\mathcal{P}}_T.$ By Gibbs' inequality \cite{mackay2003}, the exponent in Eq.~(\ref{eq:pt}) satisfies
\begin{equation}
F(\bm{r})\equiv \sum_{i=1}^{N_s} D_{\rm KL}(p^*( \cdot | \bm{r}_i - \bm{r}_{\rm s}) \parallel p(\cdot  | \bm{r}_i - \bm{r})) \ge0 ,
\end{equation}
with equality iff $p^*(h_i|\bm{r}_i-\bm{r}_{\rm s}) = p(h_i|\bm{r}_i-\bm{r})$ for each $1 \le i \le N_s$ and each $h_i$.
The set of points $\bm{r}$ where this holds depends on the number of sensors. For sufficiently large $N_s,$ this intersection should be empty unless $p^*(h_i|\bm{r}_i-\bm{r}_{\rm s}) = p(h_i|\bm{r}_i-\bm{r})$ for each $i$ and $h_i$ at $\bm{r} = \bm{r}_{\rm s}$. In other words, when the model is exact, $\mathcal{P}_T$ will remain $O(1)$ at all times, and the belief will converge to a delta-like function in the true source position. On the other hand, if the model is wrong, as it always is in our turbulent case, we can estimate the exponential long-time decay via Laplace's method:

\begin{equation}
\int_\Omega d\bm{r} \, \exp (-TF(\bm{r})) \sim A T^{-d/2}  \exp(-T F(\hat{\bm{r}})) = A T^{-d/2} \exp(-T\hat{F}) ,
\end{equation}
where $d$ is the number of spatial dimensions and $\hat{F}$ is the minimum of $F$ and is attained at the point $\hat{\bm{r}}.$ The coefficient $A$ is nonnegative and independent of $T$; it is obtained through the saddle point integration, but its precise value, related to the Hessian of the KL divergence between $p^*$ and $p,$ is irrelevant to the present discussion.

We can then conclude that for large enough $T$, $\mathcal{P}_T$ will be maximized if $p$ and $p^*$ agree at each $\bm{r}_i-\bm{r}_{\rm s},$ and it will be exponentially small if they do not. We will now show that the same conclusions also apply for $\mathcal{Z}^{(T)},$ even considering the effects of  $\mathcal{N}_{T},$ which is the key result that allows us to introduce the ranking criterion at the heart of the weighted Bayesian update (WBU) algorithm introduced in Eq.~\eqref{eq:wbu}.

In fact, each integrand appearing in $\bar{\mathcal{N}}_T$ will be sharply peaked where $D_{\rm KL}(p^*( \cdot | \bm{r}_i - \bm{r}_{\rm s}) \parallel p(\cdot  | \bm{r}_i - \bm{r'})) =0.$ For the Bernoulli distribution we consider here, this will occur on the level set of $\bm{r'}$ where $p(h_i|\bm{r}_i-\bm{r'}) = p^*(h_i|\bm{r}_i-\bm{r}_{\rm s}).$ Using the shorthand $P^*_i \equiv p^*(h_i|\bm{r}_i - \bm{r}_{\rm s}), P_i\equiv p(h_i|\bm{r}_i - \bm{r'})$ and expanding near the extremum $P_i=P_i^*+\delta P_i,$
\[
D_{\rm KL}(p^*( \cdot | \bm{r}_i - \bm{r}_{\rm s}) \parallel p(\cdot  | \bm{r}_i - \bm{r'})) = P^*_i \log P^*_i/P_i + (1-P^*_i) \log (1-P^*_i)/(1-P_i) = \frac{\delta P_i^2}{2 P^*_i (1-P^*_i)} +O(\delta P_i^3).
\]
The $i$-th integral in $\bar{\mathcal{N}}_T$ then becomes
\begin{align}
\int_\Omega d\bm{r'} \, \exp \left[-T D_{\rm KL}(p^*( \cdot | \bm{r}_i - \bm{r}_{\rm s}) \parallel p(\cdot  | \bm{r}_i - \bm{r'})) \right] & \simeq \int_\Omega d\bm{r'} \, \exp \left[-T \frac{(p(h_i|\bm{r}_i-\bm{r}')-P^*_i)^2}{2P^*_i(1-P^*_i)} \right] \nonumber \\ &\simeq \left(\frac{P^*_i(1-P^*_i)}{2\pi T}\right)^{d/2} \int_\Omega d \bm{r'} \, \delta(p(h_i|\bm{r}_i-\bm{r}') - P^*_i), \label{eq:norm}
\end{align}
where we have approximated the Gaussian by a nascent delta function. Since there is no exponential dependence on $T,$ it follows that $\mathcal{N}_T$'s contribution to $\mathcal{Z}^{(T)}$ is asymptotically sub-leading in $T,$ and therefore does not affect the result.

In summary, the utility of the overlap integral derives from being proportional to $\mathcal{P}_T,$ the posterior likelihood of the observation sequence under the given model $p,$ that is sharply maximized when the model is specified correctly. The normalization $\mathcal{N}_T,$ inherited from the sensor belief normalization, provides some numerical benefits without (in our application) meaningfully altering the asymptotic behavior of $\mathcal{Z}^{(T)}.$ Note that we explicitly used the fact that the observations are Bernoulli in order to arrive at these conclusions (specifically Eq.~(\ref{eq:norm})); while the general principle underlying WBU is robust (since $\mathcal{P}_T$ is always peaked around the correct model), the normalization we chose may have an undesirable effect for other distributions, and the precise implementation may need to be adapted. 

Given that $\cal{Z}^{(T)}$ is proportional to the likelihood of the data given the model, under an uniform prior over the models, it is also proportional to the likelihood of the model given the data. As mentioned in the main text, this means that the weights $\beta_j^{(t)}$ (Eq.~\ref{eq:wbu}) are closely related to the Akaike weights \cite{burnham2002_book}  
$w_j = {\rm Prob}(f_j| x)/\sum_i {\rm Prob}(f_i| x),$
given a set of candidate models $\{f_i\}$ and data $x.$ The Akaike weights are frequently used to assimilate inferences under multiple models.

\begin{figure}[t!]
\centering
\includegraphics[width=0.75\textwidth]{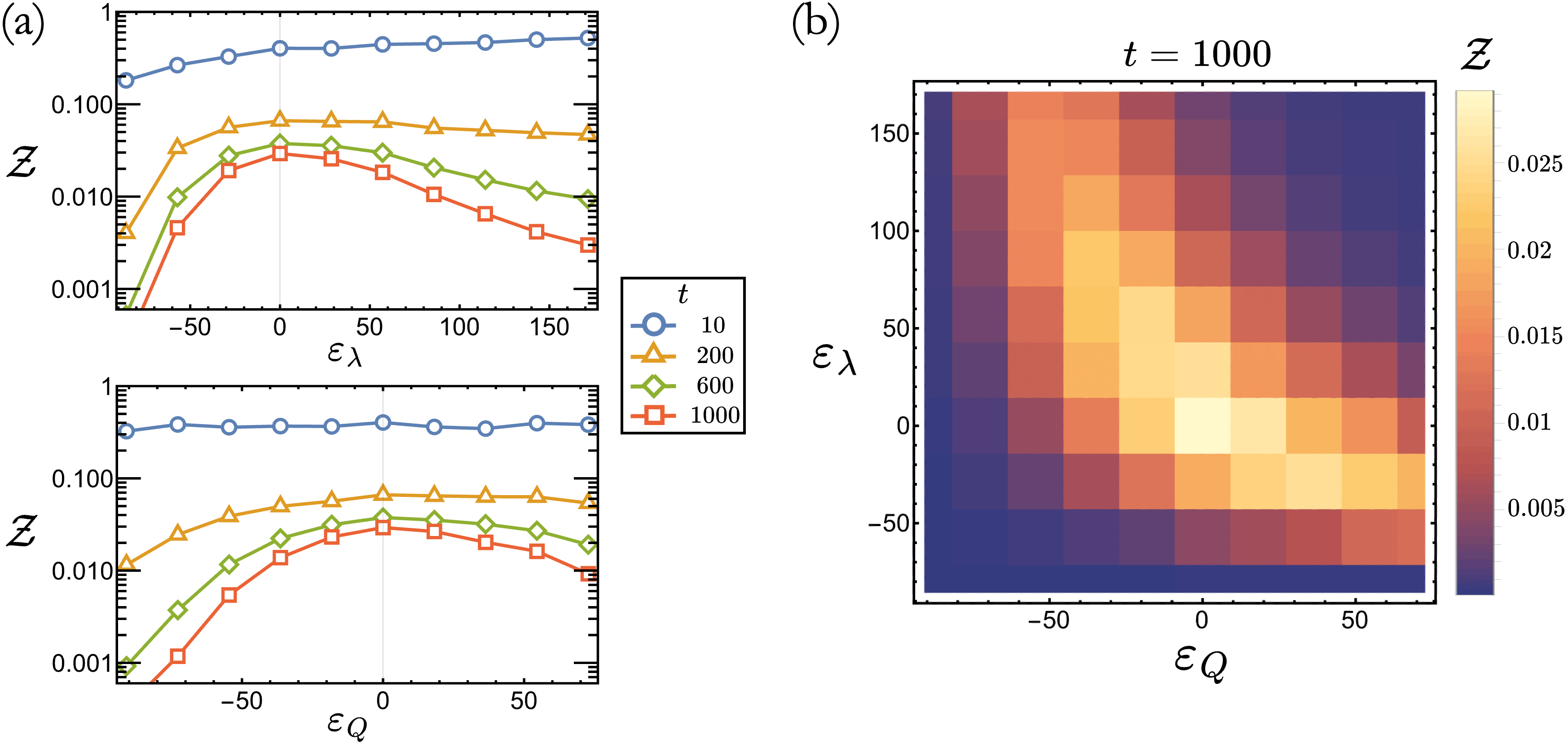}
\caption{\justifying (a) Curves of the overlap integral $\mathcal{Z}$ as a function of the percentage error on either of the model parameters. Different curves refer to distinct times, as indicated in the legend. (b) Colormap of the overlap integral $\mathcal{Z}$ at time $t=10^3$ as a function of the percentage error on both model parameters. As expected, it features a maximum when the model is exact ($\varepsilon_{\lambda}=\varepsilon_{Q}=0$) and decays to zero as we move away from it.
Data shown here have been obtained by averaging over $10^2$ episodes of time length $T=10^3$ in units of observations made by each sensor.}
\label{fig:appA}
\end{figure}

We can numerically show that the overlap integral $\mathcal{Z}$ relaxes to zero faster the further away the ground truth of the model is. To this end, let us take into account the same setup as the one shown in Fig.~\ref{fig:2}, with the source located in $\bm{r}_{\rm s}=(92,70)$ and $N_{\rm s}=16$ sensors placed on a square lattice rotated by an angle $\theta=1.04 \, \rm{rad}$ with respect to the wind direction.
We can then use the probabilistic model described in Sect.~\ref{subsec:model} to generate the observations made by the static sensors and set the source emission rate and the characteristic length scale of the turbulent odor dispersion to $Q^*=5.5$ and $\lambda^*=3.5$, respectively.
Since we know the exact model of the detection statistics, we can systematically study how the overlap integral $\mathcal{Z}$ varies depending on the error made in such model parameters. Figure~\ref{fig:appA}(a) shows the values of $\mathcal{Z}$ obtained when either the value of the characteristic length scale $\lambda$ (top panel) or the one of the emission rate $Q$ (bottom panel) is wrong. In both cases, as time progresses, the integral curve of overlap peaks more and more around the correct value of the parameter, that is, the one corresponding to a percentage error $\varepsilon_*=0$, and goes smoothly to zero as the error grows.
More generally, this still holds when both model parameters are misspecified, as shown in the colormap in Fig.~\ref{fig:appA}(b).

Let us now comment on one feature that stands out from these plots, that is, the observed asymmetry in the values of $\mathcal{Z}$ when $\lambda$ is under / overestimated. 
In fact, given the same history of measurements, the source will be estimated to be further from the sensors, the greater the estimated value of $\lambda$. This will eventually cause the overlap among the private beliefs to accumulate at the border of the finite-size arena $\Omega$ where simulations are performed. Therefore, the larger values of the overlap integral $\mathcal{Z}$ when $\lambda$ is overestimated are just finite-size effects. 
On the one hand, this is not a problem in the case of a functionally correct model, where the ground truth is attainable (as in this section) since, eventually, the correct model will still be the one with the largest value of $\mathcal{Z}$ anyway. 
However, this effect can persist in the general case of a functionally wrong model (the scenario discussed in the main text), and we are compelled to address it for consistency. 
To this end, during our analysis, we systematically set to zero the overlap values for the models that would place the source at a distance $d\leq 5\Delta x$ (with $\Delta x$ being the lattice spacing) from the boundary of the arena $\Omega$. This avoids the selection of clearly wrong models that only artificially would feature a large value of the overlap $\mathcal{Z}$, preserving the structure and the basic concepts behind the WBU approach.

\section{Details on the numerical simulations}
\label{app:num_sims}

We summarize the main parameters of the DNS in Table~\ref{tab:dnsE}. In Fig.~\ref{fig:dnsE}, we plot the turbulent energy spectrum $E(k)$ and the turbulent energy flux $\Pi(k)$ as functions of wavenumber $k$, showing the presence of an inertial range with $E(k) \propto k^{-5/3}$ and $\Pi(k) \approx \mathrm{const}$.
\begin{table}[h!]
\centering
\begin{tabular}{@{}c|c|c|c@{}}
\toprule
$N_x\times N_y\times N_z$  & $L_x\times L_y\times L_z,$ &  $dt$  & $\nu$   \\  
$1024\times 512 \times 512$ & $4\pi \times 2\pi \times 2\pi$  & $5\times 10^{-4}$  & $1.25 \times 10^{-3}$ \\ \hline
$\epsilon$    & $\tau_\eta$     & $\eta$ & $\mathrm{Re}_\lambda$\\
 $\simeq0.2$ & $\simeq0.075$ & $\simeq0.01$ & $\simeq 150$ \\
\bottomrule
\end{tabular}
\caption{\justifying Parameters of the DNS: $N_\alpha$ resolution and $L_\alpha$ dimension in direction $\alpha=x,y,z$; $dt$ time step in the DNS integration; $\nu$ kinematic viscosity; $\epsilon = \nu \langle \partial_i u_j \partial_i u_j \rangle$; $\tau_\eta = \sqrt{\nu/\epsilon}$; $\eta = (\nu^3/\epsilon)^{1/4}$; $\mathrm{Re}_\lambda = u_{\rm rms}\lambda/\nu$, where $\lambda = \sqrt{5E_{tot}/\Omega}$ is the Taylor microscale measured from the ratio between the mean system energy and enstrophy. The particles are emitted in bunches of $10^3$ particles every $10$ timesteps, and their position is stored every 150 timesteps.}
\label{tab:dnsE}
\end{table}

\begin{figure}[h!]
\centering
\includegraphics[width=0.6\textwidth]{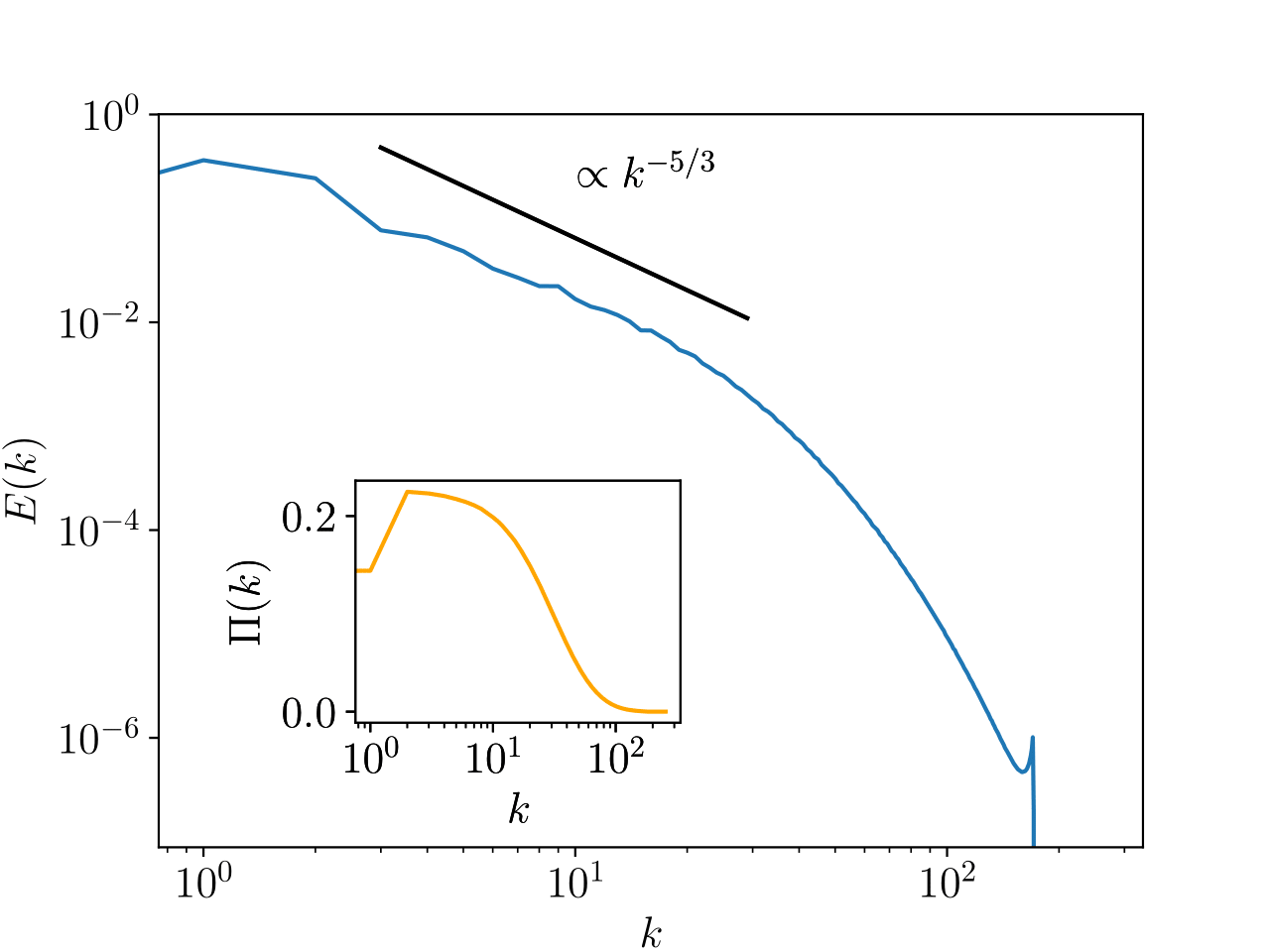}
\caption{\justifying Relevant statistical properties of our DNS: average energy spectra $E(k)$ vs $k$ and (inset) average energy flux $\Pi(k)$ vs $k$, averaged over the final $\sim 2000 \tau_\eta$ of the DNS.}
\label{fig:dnsE}
\end{figure}

In all our numerical simulations, the square arena size $\Omega$ has been set to $128\Delta x\times128\Delta x$, with $\Delta x=1$ being the lattice spacing. 
The sensors have been placed in a square grid with side length $\ell=88\Delta x$, rotated at an angle $\theta=1.04$ with respect to the mean direction of the wind.
All the data presented in Sec.~\ref{sec:results} have then been obtained by averaging over the results obtained by shifting the odor source in $30$ different locations drawn at random within an inner box $108\Delta x\times108\Delta x$ inside $\Omega$, as illustrated in Fig.~\ref{fig:appB}.
This was done to avoid boundary effects since we observed that if we place the source outside the arena or even too close to the left/right border, the belief tends to concentrate on the left/right boundary. Although it still gives us the right indication of where to look for the source, such an effect would have altered the performance statistics of the localization algorithms.

Moreover, to gather more statistics for each source location, we have divided the time series of the odor dispersion obtained from the DNS described in Sect.~\ref{subsec:setup} into $25$ runs of set time length $T=600$ (in units of the Kolmogorov timescale $\tau_\eta$).

\begin{figure}[ht!]
  \centering
  \begin{minipage}{0.45\textwidth}
    \centering
    \includegraphics[width=\textwidth]{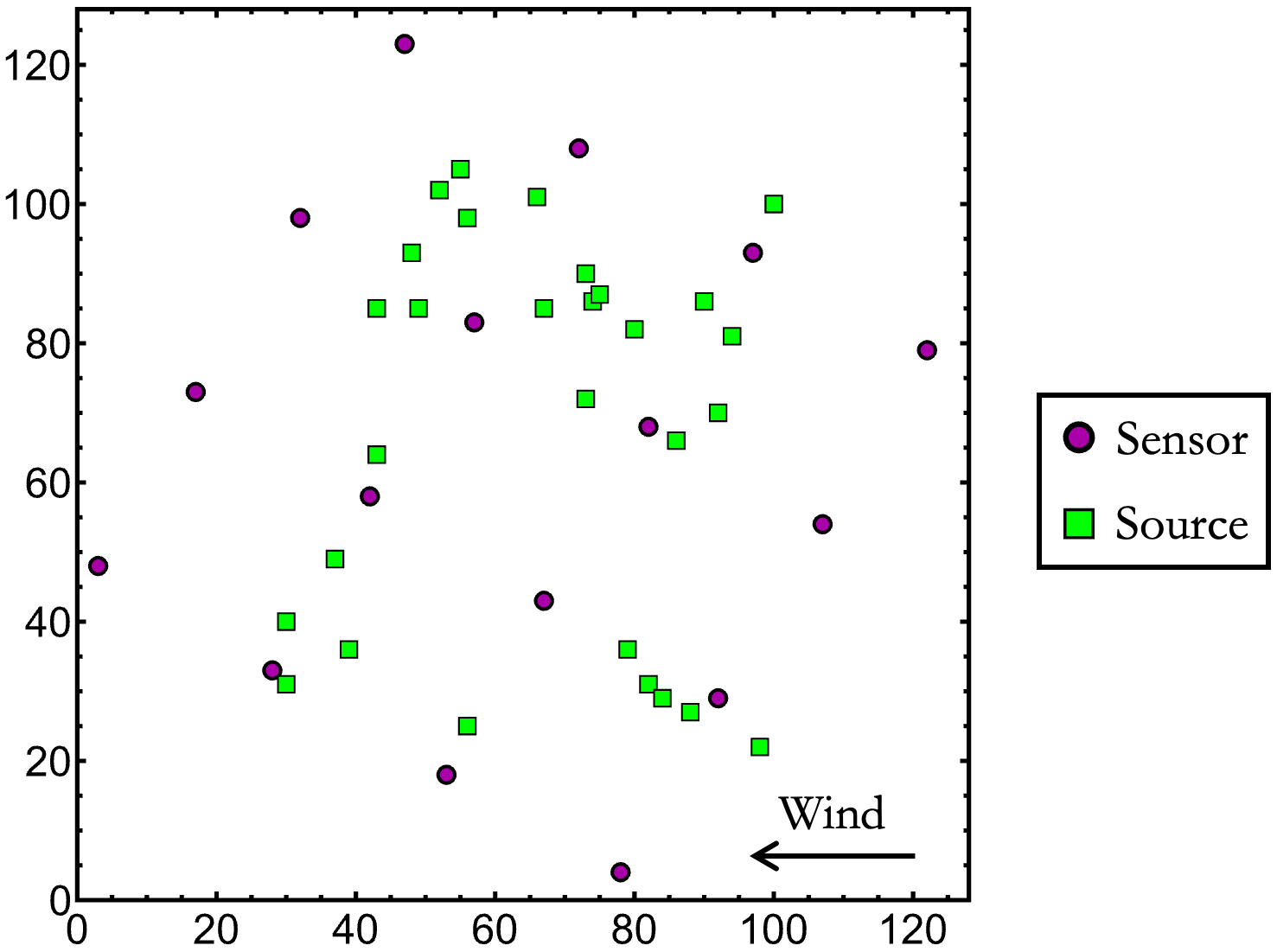}
    \caption{\justifying Illustration of the sensors' placement (magenta circles, here $N_{\rm s}=16$) and of the $30$ source locations (green squares) drawn at random inside the arena.}
    \label{fig:appB}
  \end{minipage}
  \hfill
  \begin{minipage}{0.45\textwidth}
    \centering
    \begin{tabular}{@{}ccc@{}}
\toprule
\textbf{Definition}                        & \textbf{Symbol}      & \textbf{Value} \\ \midrule
Number of samples                          & $N$                  & 50             \\
Number of MCMC perturbation steps          & $K$                  & 5               \\
Variance of proposal distribution of $\bm{r}_{\rm s}$ & $\sigma^2_{\rm pos}$ & 100            \\
Variance of proposal distribution of $\lambda$        & $\sigma^2_{\lambda}$ & 1              \\
Variance of proposal distribution of $Q$              & $\sigma^2_{Q}$       & 1      \\
Range of values of $\lambda$                 & $[\lambda_{\rm m},\lambda_{\rm M}]$  & [0,10]             \\
Range of values of $Q$                       & $[Q_{\rm m},Q_{\rm M}]$        & [0,10]  \\
\bottomrule
\end{tabular}
    \captionof{table}{\justifying Values of the hyperparameters used in the SMC algorithm.}
    \label{tab:appB}
  \end{minipage}
\end{figure}

In order to obtain the master belief in the WBU approach, we have used in each configuration a total of $M=100$ models, each of them corresponding to a different combination of parameters $\{Q,\lambda\}$ in the stochastic model introduced in Sec.~\ref{subsec:model}, with the specific values of $Q$ and $\lambda$ being $Q=\{0.7,1.7,\dots,9.7\}$ and $\lambda=\{0.7,1.7,\dots,9.7\}$, respectively.

The values of the SMC hyperparameters used in this work are summarized in Table~\ref{tab:appB}. We also checked numerically that all the results presented here are not qualitatively affected by this choice. Note that in the SMC implementation, we start from a flat prior defined over the interval $[0;10]$ for both the model parameters $Q$ and $\lambda$. Moreover, their values are bounded therein, which is, for consistency, the same range as used in the WBU approach.

Lastly, a detailed step-by-step explanation of the WBU and SMC is reported in Algorithm~\ref{alg:WBU} and Algorithm~\ref{alg:SMC}, respectively.

\begin{algorithm}[H]
\caption{Weighted Bayesian update}
\label{alg:WBU}
\begin{algorithmic}[1]
\For{$j = 1$ to $M$} \Comment{Loop over different models}
    \State Set initial common belief $B^{(0)}_j$ to a uniform distribution.
    \State Set model parameters $\bm{P}_j = \{Q_j,\lambda_j\}$.
    \For{$t = 1$ to $T$} \Comment{Loop over time}
        \State Perform sensors' measurements $\bm{h}^{(t)}$.
        \State Update common belief $B^{(t)}_j$ using Bayes' rule~\eqref{eq:bayes2}.
        \State Compute overlap integral $\mathcal{Z}^{(t)}_j$.
    \EndFor
\EndFor
\State Compute master belief $\mathcal{B}_M$ at all times from Eq.~\eqref{eq:wbu}. \Comment{Output master belief from weighted models}
\end{algorithmic}
\end{algorithm}

\begin{algorithm}[H]
\caption{Sequential Monte Carlo with Importance Sampling and perturbation step}
\label{alg:SMC}
\begin{algorithmic}[1]
\State Set initial belief $\tilde{B}^{(0)}$ to a uniform distribution.
\For{$t = 1$ to $T$}
    \State Perform sensors' measurements $\bm{h}^{(t)}$.
    \For{$i = 1$ to $N$}  \Comment{Sample's initialization}
        \State Draw sample from current belief: $\tilde{\bm{\theta}}_i \sim \tilde{B}^{(t-1)}(\bm{\theta})$.
        \State Compute weight $w_i = \mathcal{L}(\bm{h}^{(t)}|\tilde{\bm{\theta}}_i)$.
    \EndFor
    \For{$i = 1$ to $N$}
        \State Normalize weight: $w_i /= \sum_{i=1}^{N} w_i$.
        \State Compute Effective Sample Size: $N_{{\rm eff}} = 1/\sum_{i=1}^{N} w_i^2$.
    \EndFor
    \If{$N_{\text{eff}} < N_{\text{thr}}$}  \Comment{Resampling step (if needed)}
        \For{$i = 1$ to $N$}
            \State Select $\tilde{\bm{\theta}}_k$ with probability $w_k$.
            \State Put $\tilde{\bm{\Theta}}_i = \tilde{\bm{\theta}}_k$.
        \EndFor
        \For{$i = 1$ to $N$}
            \State Replace $\tilde{\bm{\theta}}_i = \tilde{\bm{\Theta}}_i$.
            \State Set uniform weights: $w_i = 1/N$.
        \EndFor
    \EndIf
    \For{$i = 1$ to $N$}  \Comment{MCMC perturbation step}
        \State Select $l\in\{1,\dots,N\}$ with probability $w_l$.
        \State Set $\bm{\theta}^{(t)}_{i,0} = \tilde{\bm{\theta}}^{(t)}_l$.
        \For{$j = 1$ to $K$}
            \State Draw new sample from proposal distribution: $\hat{\bm{\theta}}^{(t)} \sim q(\hat{\bm{\theta}}^{(t)}|\bm{\theta}^{(t)}_{i,j-1})$.
            \State Compute acceptance ratio $\alpha$ from Eq.~\eqref{eq:accept_ratio}.
            \State Draw random number $u\sim U[0,1]$.
            \If{$u < \min(1, \alpha)$}
                \State Set $\bm{\theta}^{(t)}_{i,j}=\hat{\bm{\theta}}^{(t)}$.
            \Else
                \State Set $\bm{\theta}^{(t)}_{i,j}=\bm{\theta}^{(t)}_{i,j-1}$.
            \EndIf
        \EndFor
        \State Set $\bm{\theta}^{(t)}_i = \bm{\theta}^{(t)}_{i,\tilde{B}}$.
    \EndFor
    \State Set $\tilde{B}^{(t)}(\bm{\theta}) = 1/N \sum_{i=1}^N \delta(\bm{\theta}-\bm{\theta}^{(t)}_i)$. \Comment{Output belief approximation}
\EndFor
\end{algorithmic}
\end{algorithm}

\section{Supplementary movies}
\label{app:supp_movies}

\noindent \textbf{Caption of \emph{movie$\_$dns.mp4}:} Movie showing the particles (blue) emitted from one of the five sources for the first 400 snapshots of the DNS. The source is indicated as a blue square. The thin slab over which the concentration data were coarse-grained is shown in green; particles are shown in yellow while they are within the slab.\\

\noindent \textbf{Caption of \emph{movie$\_$algorithms$\_$evolution.mp4}:} Visual comparison between the performance of the two algorithms discussed in our work, aiming to locate a source of odors (red square) advected by a turbulent flow featuring a horizontal mean wind blowing from right to left. Pink patches indicate the odor plume. The greyscale represents the probability (in logarithmic scale) of the odor source location obtained from our proposed algorithm based on Bayesian inference, i.e., the “master belief” of the weighted Bayesian update (WBU), introduced in Sec.~\ref{sec:WBU}. The green open triangles depict the candidate source positions yielded by a sequential Monte Carlo (SMC) sampling method. Both approaches use the concentration measurements made by an array of static sensors (circles) whose color codes the detection (yellow) or absence (magenta) of the odor signal. Time here is in units of observations made by each sensor, which is also equal to the Kolmogorov timescale of the flow. Note how the SMC samples strongly localize and get stuck in front of the closest sensor that detects the signal, while WBU’s master belief tends to be spread more with a tail further upwind, thus making it closer to the actual source. The two plots on the left show the time evolution of the distance d between the estimated source position and the ground truth (top panel) and of the uncertainty $\sigma_{\rm loc}$ about the source location (bottom panel) for both algorithms (gray: WBU; green: SMC) during the same episode shown on the left. Overall, this qualitative picture is consistent with the results reported in the main text.

\end{document}